\documentclass[journal,onecolumn]{IEEEtran}

\ifCLASSINFOpdf
  
\else
  
\fi

\usepackage{amsmath}
\usepackage{amsthm}
\usepackage{cases}
\usepackage{amsfonts}
\usepackage{cite}
\usepackage{amssymb}
\usepackage{enumerate}
\usepackage{graphicx}
\usepackage{float}
\usepackage{caption}
\usepackage{verbatim}
\usepackage{dblfloatfix}
\usepackage[cmintegrals]{newtxmath}
\usepackage[ruled,linesnumbered]{algorithm2e}
\usepackage{color}
\usepackage{soul}
\usepackage[acronym]{glossaries}
\newacronym{avwc}{AVWCs}{Arbitrarily Varying Wiretap Channels}
\newacronym{gavwc}{GAVWCs}{General Arbitrarily Varying Wiretap Channels}
\newacronym{ind-cpa}{IND-CPA}{indistinguishability under chosen-plaintext attack}

\newtheorem{theorem}{Theorem}
\newtheorem{lemma}{Lemma}
\newtheorem{definition}{Definition}
\newtheorem{corollary}{Corollary}
\newtheorem{remark}{Remark}

\newtheorem{example}{Example}
\begin{document}

\title{Cryptographic and Information-theoretic Security Capacities for
General Arbitrarily Varying Wiretap Channels}

\author{%
\IEEEauthorblockN{Holger Boche{$^*$}, 
Ning Cai{$^\dagger$},
    Yiqi Chen{$^*$},
    Marc Geitz{$^\ddagger$} }\\
  \IEEEauthorblockA{
{$*$} Technical University of Munich,  
80333 Munich, Germany,\;\;\{boche, yiqi.chen\}@tum.de \\
{$\dagger$} ShanghaiTech University, 201210 Shanghai, China, ningcai@shanghaitech.edu.cn\\
{$\ddagger$} T-Labs, Deutsche Telekom AG, Germany, marc.geitz@telekom.de }
\thanks{H. B. is also with the Munich Quantum Valley (MQV), Cluster of Excellence
“Centre for Tactile Internet with Human-in-the-Loop” (CeTI) and the Munich Center for Quantum Science and Technology (MCQST).

H. B. was supported by the BMBF in the programme of ”Souver\"an. Digital.
Vernetzt”, research HUB 6G-life, project identification number: 16KISK002,
and the BMBF Quantum Projects QUIET, Grant 16KISQ093, QD-CamNetz,
Grant 16KISQ077, and QuaPhySI, Grant 16KIS1598K. He also acknowledges
funding by the German Research Foundation as part of Germany’s Excellence
Strategy – EXC 2050/2 – Project ID 390696704 – Cluster of Excellence
“Centre for Tactile Internet with Human-in-the-Loop” (CeTI) of Technische
Universit\"at Dresden.

Y. C.'s work was funded by the German Research Foundation (DFG) as part of the DFG Priority Program SPP “Resilience Worlds” (SPP 2378) through the following two projects: BO 1734/42-1: “Automated Verification of Resilience and Resilience by Design” and BO 1734/45-1: “Resilience in Networked Worlds – Managing Errors, Overload, Attacks, and the Unknown.”

}}

\maketitle
\thispagestyle{empty}
\pagestyle{empty}

\begin{abstract}
We compare the strong secrecy capacities of \gls{avwc} and \gls{gavwc} with their capacities under semantic secrecy constraint and other equivalent cryptographic secrecy constraints. It turns out that the average error and strong secrecy capacity of an AVWC is always equal to its maximal error and semantic secrecy capacity. However, this equivalence does not hold for all general communication systems, and we prove this by a counterexample. We also show that, for the GAVWC, semantic security and the other cryptographic security measures considered achieve the same capacity values. Finally, we bound the gap between the strong secrecy capacity and the semantic secrecy capacity for the GAVWC. The gap vanishes if the choice of the jammer is sub-double-exponential with respect to the block length $n$, which gives a sufficient condition for the strong and semantic secrecy capacities to be equal for GAVWCs.
\end{abstract}


\section{Introduction}
This paper studies the semantic secrecy capacity of the \gls{gavwc} and its connection to the corresponding strong secrecy capacity. Semantic secrecy comes from cryptography and is considered a 'golden standard.' It is more stringent than strong secrecy, and hence the semantic secrecy capacity is always upper-bounded by the strong secrecy capacity. However, one may wonder whether the semantic secrecy capacity can be equal to the strong secrecy capacity, which is shown to be possible for weak and strong secrecy criteria \cite{maurer2000information}, and if not, how large is the gap. The goal of this work is to answer these questions.
\subsection{Literature Review}
The research on wiretap channels was started in \cite{wyner1975wire} and extended in \cite{csiszar1978broadcast}, both with the weak secrecy criterion that requires a vanishing normalized mutual information leakage given a uniform message distribution, and only some special cases were fully solved. The strong secrecy was introduced in \cite{maurer2000information}\cite{csiszar2000common} and then applied to the wiretap channel\cite{bloch2013strong} and its variations\cite{yassaee2010multiple,bjelakovic2013secrecy,goldfeld2016strong,pierrot2011strongly,schaefer2015secrecy,wang2018strong,sasaki2019wiretap,chen2025fundamental}. Compared to the weak secrecy, although still assuming a uniform message distribution, strong secrecy evaluates the information leakage by the unnormalized mutual information. 

However, although widely adopted in information-theoretic literature, the above two measures are still considered too weak to offer sufficient security to applications, as real-world messages are not always uniformly distributed. Confidential messages in the real world can be votes, any type of structured data with low entropy\cite{bellare2012semantic}, and obviously, both weak and strong secrecy cannot ensure security for these types of messages. To fulfill the requirement of a more robust and stronger security guarantee, \cite{Goldwasser1984ProbabilisticE} introduced the semantic security from a cryptographic viewpoint with the assumption that the adversary's computing power is polynomially bounded. It argued that security should hold for any message distribution and formalized the idea that extracting any information about the plaintext should be computationally infeasible.  With the semantic security guarantee, given a ciphertext, no efficient adversary should be able to extract any “useful partial information” about the plaintext beyond what was already available from side information such as message length or public context. It prohibits the adversary from computing any function of the message better than a simulator without the adversary's observation. In contemporary game-based cryptography, semantic security is most often handled via an equivalent and more operational notion: \gls{ind-cpa}. In its basic form, IND-CPA says an efficient adversary cannot distinguish encryptions of two chosen equal-length plaintexts. Under standard formalizations, semantic security and IND-CPA are equivalent (semantic security $\leftrightarrow$ IND-CPA), which is why most proofs and standards talk about IND security while still interpreting it as “semantic” privacy. The original definition of semantic security assumes an adversary who has only polynomial-bounded computing power. In \cite{bellare2012semantic}, the semantic security was introduced as an information leakage metric for wiretap channels with computationally unbounded adversaries. An analogous definition of indistinguishability was also defined in \cite{bellare2012semantic} and proved to be equivalent to semantic security with a computationally unbounded adversary. A scheme is constructed in \cite{bellare2012semantic} that achieves semantic security on binary symmetric channels. \cite{wiese2020semantic} studies security functions based on mosaic by design that achieve semantic security for memoryless discrete or Gaussian channels and for privacy amplification problems. In \cite{voichtleitner2026experimental}, an experimental validation of semantic security against a computationally unbounded adversary is studied. The work considers the distinguishing security metric and, hence, is semantically secure. In addition to classic communication scenarios, the semantic security of classic-quantum and quantum wiretap channels is studied in \cite{boche2022semantic,frey2025semantic}.

The secrecy criteria of interest in this paper are strong secrecy and semantic secrecy. In the context of secure communication over wiretap channels, semantic security (and any other equivalent forms of security aforementioned) requires vanishing information leakage under arbitrary message distributions. Related works on semantically secure communication include characterization of capacity bounds over discrete memoryless channels (DMCs) \cite{bellare2012semantic,goldfeld2016semantic,goldfeld2016arbitrarily} and practical methods \cite{ling2014semantically,campello2019semantically,wiese2020semantic,voichtleitner2026experimental}.
This work studies the communication over \gls{avwc} and the GAVWCs. It models a communication scenario in which there exists a jammer who can choose the channels for communication by his jamming strategy. Throughout this paper, we assume the jammer only knows the coding scheme, so he chooses the jamming strategy randomly, although in general we may assume that they may know more (e.g., we could assume that the jammer knows the message and/or the input of the channel). The full characterizations of the capacities of AVWC with different secrecy criteria are still open. Recent results include lower bounds of AVWC with weak secrecy \cite{molavianjazi2009arbitrary}, strong secrecy\cite{bjelakovic2013capacity,wiese2016channel,notzel2016arbitrarily,chen2021strong,chen2022strong,janda2023arbitrarily} and semantic secrecy\cite{goldfeld2016arbitrarily}. 

\subsection{Connections to Cryptography}
%In Fig. 1, the channel model is depicted using the distinguishing security, whose definition will be given later. Roughly speaking, for an arbitrary pair of messages, one of which is encoded and sent through the channel. Upon receiving the ciphertext (wiretap channel output), Eve should not be able to guess the selected message better than random guessing.

As we mentioned in the previous paragraphs, it has been proven that the semantic security and IND-CPA, in which the adversary cannot distinguish between two messages from the ciphertext better than random guessing, are equivalent under the computationally bounded attacker model.
We will later show that the equivalence between different definitions of semantic security in \cite{bellare2012semantic}\cite{Goldwasser1984ProbabilisticE} still holds when they are extended to general AVWCs and computationally unbounded adversaries.

\begin{figure}[t]
    \centering
    \includegraphics[scale=0.7]{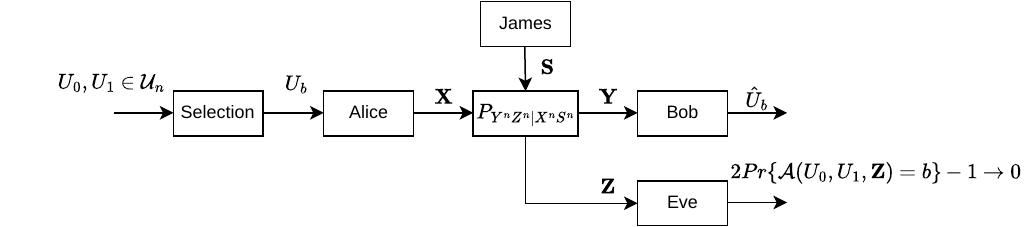}
    \caption{\footnotesize A GAVWC with distinguishing security: For an arbitrary pair of messages $(U_0,U_1)$, Eve cannot distinguish which one is sent.}
    \label{fig:model}
\end{figure}

The system model is depicted in Fig. \ref{fig:model} using the definition of distinguishing security. The semantic security constrained communication over the GAVWC can be considered as the following cryptographic game: For an arbitrary pair of messages $(U_0,U_1) \in \mathcal{U}_n$, the selector randomly picks one message from the pair with probability $(0.5,0.5)$ and transmits it. Eve observes the channel output $\textbf{Z}$ and tries to guess which message of $(U_0,U_1)$ is sent, and she succeeds if there exists an algorithm such that she can do it with a correct probability significantly larger than $1/2$. In addition, some error correcting codes are used against the intrinsic noise of the channel and the jamming signal from the malicious jammer. Hence, the systems can have at most three hostile players: The selector, James, and Eve, and four scenarios:
\begin{itemize}
    \item The selector and James are the same player;
    \item The selector and Eve are the same player;
    \item James and Eve are the same player;
    \item The selector, James, and Eve are the same player.
\end{itemize}
Our results work for the first and second cases, in which the first case corresponds to the semantic security with maximal error and average error criteria. This is because when James knows the transmitted message, he can choose a different jamming strategy for each possible message. Thus, the coding scheme must ensure the reliability and secrecy for the worst case of each message. The average error case follows as it is weaker than the maximal error criterion. For the second case, when the eavesdropper knows the message, no secrecy is possible. The latter two cases are more complicated and remain open. In fact, when James and Eve are the same player, James also has access to the wiretap channel output in a causal or strictly causal manner. This turns the problem into a completely different model. They can be considered as a cryptographic game with active adversaries, such as the Dolev-Yao attack model\cite{dolev2003security} or the tampering attackers, in which the active attacker is not only just passive eavesdropping, but also can intercept, inject, delete, reorder, and replay messages. It is a spectrum of adversary capabilities—from passive eavesdropping and traffic analysis through active manipulation (replay, Man-in-the-Middle (MITM), injection/modification/deletion/delay), and onward to endpoint compromise, side-channels, and physical tampering. Robust communication in this case has to ensure the integrity of the message so Bob can decode the message correctly, and confidentiality so Eve does not get much information about the message with the help of James. In the setting of GAVWC, James plays the role of a tampering attacker or a type of man-in-the-middle attacker who actively chooses his jamming strategy as a function of his knowledge. He alters the probability model between the sender and receivers after the information is sent to corrupt the communication between Alice and Bob and possibly help Eve to get more information about the transmitted message. 

The aforementioned attacks can substantially deteriorate the information security of a protocol. To fulfill the confidentiality and reliability requirements of the protocol, defense against advanced active attackers is always the focus of cryptographic and information-theoretic research.  Defenses in cryptography literature against active attackers include authenticated-encryption with associated-data (AEAD) \cite{rogaway2002authenticated}, authenticated key exchange (AEK)\cite{lamacchia2007stronger,diffie1992authentication} or hardware and operational controls. On the other hand, in information-theoretic literature, additional resources such as randomness, secret keys that are only available between Alice and Bob, are used to communicate against jamming and eavesdropping.

\subsection{Main Contributions}

We are interested in the comparison between (general) AVWCs' strong secrecy capacity and (general) AVWCs' semantic secrecy capacity. Consequently, there are three natural questions:
\begin{itemize}
    \item Whether for an AVWC, the capacity of semantically secure codes is always equal to the capacity of strongly secure codes;
    \item Whether for all communication systems, semantically secure codes and strongly secure codes have the same capacity;
    \item How large the gap between the capacities of semantically secure codes and strongly secure codes can be in a communication system if the size of the system is fixed.
\end{itemize}
As semantic secrecy is a stronger criterion than strong secrecy, the strong secrecy capacity is always lower bounded by the semantic secrecy capacity. However, one may not prove the first question directly by showing the achievability of the capacity of strongly secure codes by semantically secure codes, because the (single-letter) capacity of strongly secure codes is still unknown. If the answer to the first question was positive, the second question would be raised, and if the answer to the second question is negative, one might want to know how large the gap is, which raises the third question. Specifically, we show that for an arbitrarily varying wiretap channel, its maximal error and semantic secrecy capacity is equal to its average error and strong secrecy capacity. However, this does not hold for all communication systems. We give a negative answer to the second question by providing an example, in which the system has a positive strong secrecy capacity but zero semantic secrecy capacity. Lastly, we answer the third question by showing that the gap between the capacities of semantically secure code and strongly secure code is upper-bounded by the double logarithm of the number of jamming strategies. Consequently, the semantic secrecy capacity is always equal to the strong secrecy capacity if James has only sub-double-exponential number of jamming strategies.\footnote{Throughout this paper, a random variable, its sample value, and its alphabet are denoted by a capital letter $X$, lowercase letter $x$, and caligraphic letter $\mathcal{X}$, respectively. A random $n-$length sequence and its sample value are denoted by boldface symbols ${\bf X}$ and ${\bf x}$. The cardinality of a set $\mathcal{X}$ is written by $|\mathcal{X}|$.} 

\section{Definitions}\label{sec: definitions}
A general arbitrarily varying wiretap channel (GAVWC) $({\cal W}^{G}, {\cal V}^{G})$ consists of a sequence of pairs $(\mathcal{W}^{G,n}, \mathcal{V}^{G,n})$, of sets of channels for $n=1,2,\ldots $. We assume that the channel GAVWC is  controlled by an adversary, James, who may arbitrarily choose a pair  $(W^{G,n}, V^{G,n}) $ from $({\cal W}^{G,n}, {\cal V}^{G,n})$ for each $n$, where we assume that for each $n$ and all  $(W^{G,n}, V^{G,n}) ,(W^{G,n'}, V^{G,n'}) \in ({\cal W}^{G,n}, {\cal V}^{G,n})$, $W^{G,n}, V^{G,n},W^{G,n'}, V^{G,n'}$ have the same input alphabet ${\cal X}^{G,n}$, the output alphabets of $W^{G,n}$ and $W^{G,n'}$ are the same set $\mathcal{Y}^{G,n}$,  the output alphabets of $V^{G,n}$ and $V^{G,n'}$ are the same set $\mathcal{Z}^{G,n}$, but possibly the output alphabets of  $W^{G,n}$ and $ V^{G,n}$ are different. To define the transmission rate and capacity properly, we always assume that the cardinalities of the input and two output alphabets are exponentially increasing as $n$ increases. Hence, each of the alphabets $\mathcal{X}^{G,n}, \mathcal{Y}^{G,n}$ and $\mathcal{Z}^{G,n}$ is the  Cartesian product of $n$ finite sets, which are not necessarily the same. For each choice  $(W^{G,n}, V^{G,n})  \in ({\cal W}^{G,n}, {\cal V}^{G,n})$ by the jammer, the common input of  $W^{G,n}$ and $ V^{G,n} $ is accessed by a legal sender, and a legal receiver and the second  adversary, the eavesdropper, access the outputs of $W^{G,n}$ and $V^{G,n}$ respectively. The arbitrarily varying wiretap channel $(\mathcal{W},\mathcal{V})$ is a special case of the GAVWC where the channel is discrete memoryless and for each $n$ the alphabet $\mathcal{X}^n$ is the Cartesian product of the same alphabet $\mathcal{X}$. The two output alphabets follow similarly. In this case, $\mathcal{W}$ and $\mathcal{V}$ are two finite sets of DMCs $\{W_s:s\in\mathcal{S}\}$ and $\{V_s:s\in\mathcal{S}\}$. We call $s\in\mathcal{S}$ the state of the channel. The sender wants to send a message to the legal receiver reliably and keep the message in secret from the eavesdropper. Similar to AVC or AVWC, by considering the worst case, we assume that the jammer chooses the worst pair $(W^{G,n},V^{G,n})$ for the legal communicators (sender and receiver), the eavesdropper knows pair $(W^{G,n},V^{G,n})$, and neither the sender nor the receiver knows the pair. For each $n$,  the channels in the sets  ${\cal W}^{G,n}$ and ${\cal V}^{G,n}$ are called main channels and wiretap channels, respectively.

In the following, we give the definition of a random encoding code for the GAVWC.
\begin{definition}
    A random encoding code $({\Phi}_n, {\psi}_n)$ of length $n$  for a GAVWC $({\cal W}^{G,n}, {\cal V}^{G,n})$  consists of
    \begin{itemize}
        \item A message set ${\cal U}^{G,n}$;
        \item A random encoder ${\Phi}_n:{\cal U}^{G,n} \to {\cal X}^{G,n}$;
        \item A deterministic decoder $\phi_n: {\cal Y}^{G,n} \to {\cal U}^{G,n}$,
    \end{itemize}
    where the random encoder $\Phi$ is a random variable distributed on a set of encoders $\{\phi\}$, where each $\phi:{\cal U}^{G,n} \to {\cal X}^{G,n}$ is a conditional probability distribution on ${\cal X}^{G,n}$.
\end{definition}
Thus, $W^{G,n}$ and $V^{G,n}$ produce outputs ${\bf y}$ and ${\bf z}$ with probabilities $W^{G,n}({\bf y}|\Phi_n(u)):=\sum_{\bf x} \Phi_n({\bf x}|u)W^{G,n}({\bf y}|{\bf x})$ and  $V^{G,n}({\bf z}|\Phi_n(u)):=\sum_{\bf x} \Phi_n({\bf x}|u)V^{G,n}({\bf z}|{\bf x})$, respectively, when a message $u \in {\cal U}^{G,n}$ is encoded by $\Phi$ and sent to the channel, if  the jammer chooses $(W^{G,n}, V^{G,n})  \in ({\cal W}^{G,n}, {\cal V}^{G,n})$ to disturb the transmission.

In the following, we give brief descriptions of some well-known information leakage measures for secure communication over normal discrete memoryless wiretap channels (without varying channel states). The formal definitions of achievable rate with these security measures of GAVWCs will then be provided. Let $\textbf{Z}$ be the output of the wiretap channel. The first information leakage measure used for the wiretap channel is the \emph{weak secrecy}:
\begin{align}
    \lim_{n\to\infty}\frac{1}{n} I(U;\textbf{Z})= 0,
\end{align}
where the message random variable $U$ is uniformly distributed. Enhanced in \cite{maurer2000information,csiszar2000common}, the strong secrecy is defined by
\begin{align}
    \lim_{n\to\infty}I(U;\textbf{Z})= 0,
\end{align}
where the message random variable $U$ is uniformly distributed. The only difference is that the normalized information leakage is replaced with a whole block of information leakage. Semantic security is a concept from the cryptography community and was extended to the wiretap channel setting in \cite{bellare2012semantic} as the advantage of a computationally unbounded adversary for computing a function of the message when the ciphertext (the wiretap channel output in our case) is available:
\begin{align}\label{def: advantage 1}
    \max_{P_U,f} \left(\max_{\mathcal{A}} Pr\left\{\mathcal{A}(\mathbf{Z}(\mathrm{Enc}(U)))=f(U)\right\} - \max_{\mathrm{S}}Pr\left\{\mathrm{S}(\mathcal{U}^{G,n}) = f(U)\right\}\right)\to 0,
\end{align}
where $f$ is a function of the message, $\mathbf{Z}(\mathrm{Enc}(U))$ is the wiretap channel output when the mesage $U$ is transmitted using the encoding function $\mathrm{Enc}$, $\mathcal{A}$ is the algorithm of the eavesdropper who observes $\mathbf{Z}$, $\mathrm{S}$ is a simulator and $\mathcal{U}^{G,n}$ is the message set. Furthermore, the \emph{distinguishing advantage} is defined in \cite{bellare2012semantic} by
\begin{align}
    \label{def: advantage 2}&\max_{u_0,u_1,\mathcal{A}} 2Pr\{\mathcal{A}(u_0,u_1,\textbf{Z}(u_b))=b\}-1\\
    &=\max_{u_0,u_1} \mathrm{SD}(\textbf{Z}(u_0),\textbf{Z}(u_1))\to 0,
\end{align}
where $b\in\{0,1\}$ is uniformly distributed, $\mathcal{A}$ is any algorithm of the adversary and $\mathrm{SD}$ is the total variational distance.

The above two definitions are not widely used in information-theoretic literature, as we introduced before, where information leakage is usually measured by (normalized) mutual information. \cite{bellare2012semantic} further translates the measures \eqref{def: advantage 1}, \eqref{def: advantage 2} using the language of information theory, and 
shows the equivalence between \eqref{def: advantage 1}, \eqref{def: advantage 2} and the following mutual information constraint:
\begin{align}
    \lim_{n\to \infty} \max_{P_U}I(U;\mathbf{Z})\to 0,
\end{align}
which is called the \emph{mutual information security} in \cite{bellare2012semantic}.
Denote by 
%and $\max_{V_n \in {\cal V}_n}I(U;{\bf Z}(V_n)) \rightarrow 0$ as $n \rightarrow \infty$
\begin{equation} \label{eq_D}
D_{V^{G,n}}(u):=\sum_{\bf z} V^{G,n}({\bf z}|\Phi_n(u))\log \frac{V^{G,n}({\bf z}|\Phi_n(u))}{\sum_{v}\frac{1}{|{\cal U}^{G,n}|}V^{G,n}({\bf z}|\Phi_n(v))},
\end{equation}
and
\begin{equation} \label{eq_W}
W^{G,n}(\psi_n({\bf Y})\not=u|\Phi_n(u)):=\sum_{{\bf y}: \psi({\bf y})\not=u} W^{G,n}({\bf y}|\Phi_n(u)).
\end{equation}
We can analogously define semantic security and distinguishing security for general AVWCs:
\begin{align}
    &\mathrm{Adv}^{SS}(\mathcal{V}^{G,n}) :=\max_{P_U,f,V^{G,n}} \left(\max_{\mathcal{A}} Pr\left\{\mathcal{A}(\mathbf{Z}(V^{G,n},\mathrm{Enc}(U)))=f(U)\right\} - \max_{\mathrm{S}}Pr\left\{\mathrm{S}(\mathcal{U}^{G,n}) = f(U)\right\}\right),\\
    &\mathrm{Adv}^{DS}(\mathcal{V}^{G,n}) := \max_{\mathcal{A},u_0,u_1,V^{G,n}} 2Pr\{\mathcal{A}(u_0,u_1,\textbf{Z}(V^{G,n},u_b))=b\}-1\\
    &\quad\quad\quad\quad\quad=\max_{u_0,u_1,V^{G,n}} \mathrm{SD}(\textbf{Z}(u_0,V^{G,n}),\textbf{Z}(u_1,V^{G,n})).
\end{align}
\begin{remark}\label{rem: safty margins}
    In our communication scenario involving the wiretap channel with a jammer, the security measures immediately make the influence of the individual unauthorized parties—wiretapper Eve, jammer James, and the selector—analytically clear. The selector chooses the message pair or an input distribution. However, Eve could also do the same before the transmission begins. Additionally, the selector can also choose a function f in the semantic security scenario; Eve can make this choice. Furthermore, James selects the jamming strategy that is best for him. All three players attempt to achieve the maximum of their respective security measures. These are all deterministic optimization problems, and deterministic strategies exist for each player. Given the structure of the security functions, it can be assumed that the selector, James, and Eve know the optimal strategies of all players.
\end{remark}

 In the following, we provide definitions for achievable codes with different security criteria.
\begin{definition}
    A non-negative $R$ is achievable for a GAVWC by random encoding codes  with average probability of error and strong secrecy criteria (or an average error and strongly secure codes in short), if for all positive $\epsilon, \lambda,\mu$ and all sufficiently large $n$, there exists a  random encoding code $({\Phi}_n, {\psi}_n)$ of length $n$ with a rate larger than $R-\epsilon$ such that  for uniform random message $U$ on ${\cal U}^{G,n}$ and all channel $W^{G,n} \in {\cal W}^{G,n}$
 the average probability of error 
\begin{equation} \label{eq_ex01}
\sum_{u \in {\cal U}^{G,n}}\frac{1}{|{\cal U}^{G,n}|} W^{G,n}(\psi_n({\bf Y})\not=u|\Phi_n(u)) \le  \lambda
\end{equation}
and the mutual information for uniform random message $U$ on ${\cal U}^{G,n}$ and for all $V^{G,n} \in {\cal V}^{G,n}$, leaking to the eavesdropper
\begin{equation} \label{eq_ex02}
I(U;{\bf Z}(V^{G,n}))= \sum_{u \in {\cal U}^{G,n}}  \frac{1}{|{\cal U}^{G,n}|}D_{V^{G,n}}(u) \le \mu,
%I(U;{\bf Z}(V_n))= \sum_{u \in {\cal U}_n}  D_{V_n}(u) \log \frac{D_{V_n}(u)}{\sum_{v \in {\cal U}_n}D_{V_n}(v)   } \le \mu_n   \rightarrow 0
\end{equation}
% as $n \rightarrow \infty$, 
 where ${\bf Z}(V^{G,n})$ is the random output of $V^{G,n}$ for the uniform random message $U$ by applying the random encoder $\Phi_n$. The capacity of GAVWC $({\cal W}^{G,n},{\cal V}^{G,n})$ for average error and strongly secure codes is the maximum achievable rate of the code, denoted by $C_{a-str}({\cal W}^{G,n},{\cal V}^{G,n})$.
\end{definition}
\begin{remark}\label{rem: average case}
    In \eqref{eq_ex01} and \eqref{eq_ex02}, the impact of the choice of jammer becomes clear, with two performance metrics now being influenced by the jammer. The same observations apply to \eqref{eq_ex01} and \eqref{eq_ex02} as those discussed in Remark \ref{rem: safty margins}.
\end{remark}
Moreover, for a technical reason, which we shall see soon, we assume $D_{V^{G,n}}(u)$ is uniformly upper bounded by a real sequence $\{\delta_n\}$, i.e., 
\begin{equation} \label{eq_ex02a}
D_{V^{G,n}}(u)\le \delta_n
\end{equation}
 for all $u \in {\cal U}^{G,n}$ and all $V^{G,n} \in {\cal V}^{G,n}$, and all $n$.

\begin{definition}
    A code satisfies the maximum error  criterion if (\ref{eq_ex01}) is replaced by 
%\begin{equation} \label{eq_ex01a}
% W_n(\psi({\bf Y})\not=u|\Phi(u)) \le  \lambda_n \rightarrow 0
%\end{equation}
\begin{equation} \label{eq_ex01a}
 W^{G,n}(\psi_n({\bf Y})\not=u|\Phi_n(u)) \le  \lambda
\end{equation}
for all $u \in {\cal U}^{G,n}$ and all  channel $W^{G,n} \in {\cal W}^{G,n}$ and a code satisfies the  semantically secure criterion if (\ref{eq_ex02}) holds for an arbitrarily distributed random message $U$.
A  non-negative $R$ is achievable for a GAVWC by random encoding code  with maximum probability of error and mutual-information secure (or a maximum error and MI secure code in short) if for all $\epsilon,\lambda, \mu >0$ and  sufficiently large $n$ there exists a code of length $n$ with a rate larger than $R-\epsilon$, satisfying both maximum error  criterion and semantically secure criterion for $\lambda$ and $\mu$. The capacity of GAVWC $({\cal W}^{G,n},{\cal V}^{G,n})$ for maximum error and semantically secure codes is the maximum achievable rate of the code, denoted by $C_{m-mis}({\cal W}^{G,n},{\cal V}^{G,n})$. 

Similarly, when the secure criterion of the code is replaced with 
\begin{align}
    \mathrm{Adv}^{SS}(\mathcal{V}^{G,n}) \leq \mu,
\end{align}
we say it is a maximum error and semantically secure code, and denote the maximum achievable rate of the code by $C_{m-sem}({\cal W}^{G,n},{\cal V}^{G,n})$.

Lastly, when the secure criterion of the code is 
\begin{align}
    \mathrm{Adv}^{DS}(\mathcal{V}^{G,n}) \leq \mu,
\end{align}
it is a maximum error and indistinguishable code, and the maximum achievable rate of the code is denoted by $C_{m-ind}({\cal W}^{G,n},{\cal V}^{G,n})$.
\end{definition}
\begin{remark}
    The same observations apply to the performance functions introduced in the definition and the corresponding capacities as in Remark \ref{rem: average case}.
\end{remark}

\begin{lemma}\label{lem: equivalence}
    For GAVWCs, the following three statements are equivalent:
    \begin{itemize}
        \item A scheme for GAVWCs achieves semantic security, i.e., $\mathrm{Adv}^{SS}(\mathcal{V}^{G,n}) \to 0$;
        \item A scheme for GAVWCs achieves distinguishing security, i.e., $\mathrm{Adv}^{DS}(\mathcal{V}^{G,n})\to 0$;
        \item A scheme for GAVWCs achieves mutual information security, i.e., $\lim_{n\to \infty} \max_{P_U}I(U;\mathbf{Z}) \to 0$.
    \end{itemize}
\end{lemma}
The proof is almost the same as the proof in \cite{bellare2012cryptographic}. We provide it in Appendix \ref{sec: proof of lem equivalence} for completeness. We show that each of the adversary's advantages can be upper-bounded by a function of another advantage that vanishes as the block length goes to infinity.

The notation of semantic secrecy (and its equivalent definitions) is considered the 'gold standard' in cryptography. It implies that no adversary (computationally unbounded in information-theoretical security) can exploit whatever information does exist (if any) to compute any predicate/function of $U$ better than a simulator.
Obviously, as semantic secrecy is stronger than strong secrecy, $C_{a-str}({\cal W}^{G,n},{\cal V}^{G,n})\ge C_{m-mis}({\cal W}^{G,n},{\cal V}^{G,n})=C_{m-sem}({\cal W}^{G,n},{\cal V}^{G,n})=C_{m-ind}({\cal W}^{G,n},{\cal V}^{G,n})$ always holds.

We adopt asymmetric settings for Bob and Eve for all security measures. When semantic security is required at Eve, one may replace the maximal error with the inverse of semantic security: For any function $f$ on the message set, there exists an algorithm $\mathcal{A}$ such that Bob can recover the value of the function:
\begin{align}
    \max_{P_U,W^{G,n}}  Pr\left\{\mathcal{A}(\mathbf{Y}(W^{G,n},\mathrm{Enc}(U)))\neq f(U)\right\} \leq \lambda.
\end{align}
It is straightforward to verify that this requirement is equivalent to the maximal error criterion, as the function $f$ can be set to the identity function. On the other hand, when distinguishing security is the security measure, one can define a code as distinguishable at Bob if
\begin{align}
    \max_{u_0,u_1,W^{G,n}}Pr\{\mathcal{A}(u_0,u_1,\textbf{Y}(W^{G,n},u_b))\neq b\} \leq \lambda.
\end{align}
If we denote the capacity for this case by $C_{d-ind}({\cal W}^{G,n},{\cal V}^{G,n}),$ it follows that $C_{m-ind}({\cal W}^{G,n},{\cal V}^{G,n})\leq C_{d-ind}({\cal W}^{G,n},{\cal V}^{G,n})$ as distinguishability is weaker than the maximal error criterion. One sufficient condition for them to be equal is that $\lambda$ decreases double-exponentially fast as $n$ grows. Then we can extract a maximal error code by removing half of the bad messages from the code that achieves $C_{d-ind}$ and construct the maximal error decoder by defining the decoding set for each message $u_i$ as
\begin{align}
    \mathcal{D}(u_i) = \{\textbf{y}:\mathcal{A}(u_i,u_j,\textbf{y})=i,\forall j\neq i\}.
\end{align}
The decoding error is $(|\mathcal{U}^{G,n}|-1)\lambda\to 0$ when $\lambda$ is double-exponentially small. Hence, using the maximal error criterion for Bob is a worst-case consideration that assumes Bob always knows nothing, but Eve knows that the transmitted message can only be one of the messages $(u_0,u_1).$

It is worth revisiting the remarks on Definitions 2 and 3. James uses jamming strategies to interfere with the respective channels from Alice to Bob and from Alice to Eve. Through these jamming strategies, he also affects the respective performance metrics for these two channels. For a whole range of practically relevant channels, however, James cannot attack these performance metrics independently of one another. This occurs, for example, when James can employ multi-antenna systems, MIMO signal processing, and beamforming. In such cases, there is generally a trade-off between the interference of the two performance functions. In this work, however, the practically important assumption is made that Alice and Bob have no knowledge of the respective strategies of the respective attackers. The two performance metrics must be satisfied for all possible attacker strategies through appropriate coding. For James, this requirement implies that the corresponding maximizations over the channels in \eqref{eq_ex01} and \eqref{eq_ex02} can be performed independently of one another. For beamforming or MIMO signal processing, this implies that the same performance can only be achieved as in the case where James can attack the corresponding channels using independent beamforming strategies.

Another important property of AVCs, symmetrizability\cite{csiszar1988capacity}\cite{ericson1985exponential}, determines the positivity of the reliable communication rate of AVCs, which is defined as follows:
\begin{definition}
    An AVC is $\mathcal{X}-$symmetrizable if there exists a transition matrix $T:\mathcal{X}\to \mathcal{S}$ such that
    \begin{align}
        \sum_{s\in\mathcal{S}}W_s(y|x)T(s|x') = \sum_{s\in\mathcal{S}}W_s(y|x')T(s|x)
    \end{align}
    for all $x,x'\in\mathcal{X}$ and $y\in\mathcal{Y}$.
\end{definition}
It was proved in \cite{csiszar1988capacity} that the capacity of an AVC without common randomness at Alice and Bob is positive if and only if the channel is not $\mathcal{X}-$symmetrizable. For a $\mathcal{X}-$symmetrizable AVC, James can choose a codeword $\textbf{x}$ from the codebook and generate the jamming sequence i.i.d. according to the distribution $T$. By the symmetrizable condition, the jamming sequence makes any other codewords in the codebook indistinguishable to the decoder compared to \textbf{x}. The symmetrizability causes the difference between the deterministic code capacity and the common randomness-assisted code capacity. In fact, \cite{ahlswede1978elimination} further shows that the deterministic code capacity of an AVC is either 0 or equal to its common randomness-assisted code capacity. Obviously, secure communication is not possible over a symmetrizable AVC without common randomness assistance, as reliable communication is not feasible.

\section{main results}
This section provides the main results of this paper. We answer the three questions proposed in the Introduction in the order that they are posed.

In the following, we shall use the sequences of  $\frac{1}{n} \log \log |{\cal W}^{G,n}|$,  $\frac{1}{n} \log \log |{\cal V}^{G,n}|$, and $\frac{1}{n} \log \log v_n^{-1}$   in the following. Rigorously, the case where $|{\cal W}^{G,n}|$, $|{\cal V}^{G,n}|$ or $v_n$ is one, should be excluded and discussed separately, to avoid the undefined notation ``$\log 0$". But for convenience of notation, we shall not distinguish them, but understand $ \log \log |{\cal W}^{G,n}|$, $\log \log |{\cal V}^{G,n}|$ or $\log \log v_n^{-1}$, as $0$ in the case. It is easy to see that this will make no problem  in the statements and proofs below.
\begin{theorem}\label{thmavwc}
The capacities of AVWCs and GAVWCs satisfy
\begin{enumerate}
    \item For all AVWCs $({\cal W}, {\cal V})$, $C_{a-str}({\cal W},{\cal V})=C_{m-mis}({\cal W},{\cal V})$. In detail, if $\mathcal{W}$ is $\mathcal{X}-$symmetrizable and no common randomness is available at Alice and Bob,
    \begin{align}
        C_{a-str}({\cal W},{\cal V})=C_{m-mis}({\cal W},{\cal V}) = 0.
    \end{align}
    If $\mathcal{W}$ is not $\mathcal{X}-$symmetrizable or there is common randomness available at Alice and Bob,
    \begin{align}
        C_{a-str}({\cal W},{\cal V})=C_{m-mis}({\cal W},{\cal V}) = \lim_{n\to\infty}\frac{1}{n}\max_{P_U,P_{\textbf{X}|U}}\left(\min_{q\in\mathcal{P}(\mathcal{S})} I(U;\textbf{Y}) - \max_{s^n\in\mathcal{S}^n} I(U;\textbf{Z}|s^n) \right),
    \end{align}
    where $\mathcal{P}(\mathcal{S})$ is the set of distributions on the state alphabet $\mathcal{S}$, the joint distribution of $(U,\textbf{X},\textbf{Y},\textbf{Z})$ given $\textbf{S}$ is
    \begin{align}
        P_UP_{\textbf{X}|U}\left(\prod_{i=1}^n \sum_{s\in\mathcal{S}} q(s)W_{s_i}(y_i|x_i) \right)V^n_{\textbf{s}}(\textbf{z}|\textbf{x}).
    \end{align}
    \item More generally, for a GAVWC $({\cal W}^{G,n}, {\cal V}^{G,n})$, $C_{a-str}({\cal W}^{G,n},{\cal V}^{G,n})=C_{m-mis}({\cal W}^{G,n},{\cal V}^{G,n})$, if 
\begin{equation} \label{eq_thmAVC}
\begin{aligned}
    \lim_{n \rightarrow \infty} \frac{1}{n} \log \log |{\cal W}^{G,n}|&=\lim_{n \rightarrow =\infty} \frac{1}{n} \log \log |{\cal V}^{G,n}|=\lim_{n \rightarrow \infty} \frac{1}{n} \log \log v_n^{-1}=0
\end{aligned}
\end{equation}
where $v_n$ is the minimum value of $V^{G,n}({\bf z}|{\bf x})$ over all $V^{G,n}, {\bf x}$ and ${\bf z}$ with $V^{G,n}({\bf z}|{\bf x})>0$.
\end{enumerate}
\end{theorem}

The proof of Theorem \ref{thmavwc} heavily relies on Lemma \ref{lemma_semaim}, which is provided at the end of this section, and its proof is given in Appendix \ref{sec: proof of the thm}. The regularized capacity formula of AVWCs, i.e., the representation of the capacity as the limit of a sequence of corresponding n-letter expressions, was provided in \cite{wiese2016channel}\cite{notzel2016arbitrarily}. When mutual information security and maximum error criteria are applied, we consider a case that is worse than average error and strongly secure, in which the eavesdropper can freely decide the distribution of the message. Intuitively, it seems to us that the eavesdropper would be at a very advantageous position if we use a semantically secure criterion. Surprisingly, our results show that the eavesdropper may not take any advantage in rate at all, unless the number of his choices double exponentially increases or faster as $n$ grows. Hence, we would not need to care whether the secure criterion is strong or semantic if the choices of the eavesdropper are sub-double-exponential. Moreover, the following result shows that replacing the mutual information security with semantic security or distinguishing security, Theorem \ref{thmavwc} still holds. 

\begin{corollary}
For GAVWCs,
\begin{align}
    C_{m-mis}({\cal W}^{G,n},{\cal V}^{G,n}) = C_{m-sem}({\cal W}^{G,n},{\cal V}^{G,n}) = C_{m-ind}({\cal W}^{G,n},{\cal V}^{G,n}).
\end{align}
    Consequently, the statements in Theorem \ref{thmavwc} still hold if $C_{m-mis}({\cal W},{\cal V}) (C_{m-mis}({\cal W}^{G,n},{\cal V}^{G,n}))$ is replaced with $C_{m-sem}({\cal W},{\cal V})$ $(C_{m-sem}({\cal W}^{G,n},{\cal V}^{G,n}))$ or $C_{m-ind}({\cal W},{\cal V})(C_{m-ind}({\cal W}^{G,n},{\cal V}^{G,n}))$.
\end{corollary}
The corollary directly follows from Theorem \ref{thmavwc} and Lemma \ref{lem: equivalence}.

The next theorem answers the second question: The semantic secrecy capacity of a communication system can be smaller than the strong secrecy capacity.
\begin{theorem}\label{the: question 2}
    Generally, the semantically secure capacity of a communication system is smaller than or equal to its strongly secure capacity, and the equality does not always hold.
\end{theorem}
We prove the theorem by constructing a counterexample in Appendix \ref{sec: proof of theorem question 2} that shows that
\begin{align}
        C_{a-str}({\cal W}^{G,n},{\cal V}^{G,n})> C_{m-mis}({\cal W}^{G,n},{\cal V}^{G,n}).
\end{align}
In fact, we show that the semantically secure rate can be 0 when the strongly secure rate is positive, and this is still true even when the secrecy criterion is relaxed to weak mutual information secrecy, i.e., $\lim_{n\to \infty}\max_{P_U,V^{G,n}} \frac{1}{n}I(U,\textbf{Z}(V^{G,n}))\to 0.$

So far, we have had a negative answer to the second question: there exists a communication system for which, the capacities of semantically secure codes and strongly secure codes are different. In the constructed example, the eavesdropper who has double exponential choices, as $n$ increases, is too powerful. Indeed, since ${\cal V}^{G,n}$ has size \[|{\cal V}^{G,n}|={ 2^n \choose 2^{na}}=\frac{\prod_{i=0}^{2^{na}-1} [2^n-i]}{(2^{na})!},\]
we have that
\[(2^{n(1-a)}-1)^{2^{na}}=\frac{(2^n-2^{na})^{2^{na}}}{(2^{na})^{2^{na}}}<\frac{(2^n-2^{na})^{2^{na}}}{(2^{na})!}<|{\cal V}^{G,n}|<(2^n)^{2^{na}},\]
which yields \[ \lim_{n \rightarrow \infty}\frac{1}{n} \log \log |{\cal V}^{G,n}|=a.\]
Naturally, one might ask whether there a such an example with a smaller size that exists or how large the gap between  the capacities of semantically secure codes and strongly secure codes can be in a communication system, like in the proof of Theorem \ref{the: question 2}, if the size of the system is fixed, which is the third question we proposed in the introduction, and is answered by the following theorem.

\begin{theorem} \label{the: question 3}
For all GAVWC $({\cal W}^{G,n},{\cal V}^{G,n})$ with  
\begin{equation} \label{eq_vn}
\limsup_{n \rightarrow \infty} \frac{1}{n} \log \log v_n^{-1}=0,
\end{equation}
\begin{eqnarray}\label{eq_thm2}
&&C_{a-str}({\cal W}^{G,n},{\cal V}^{G,n})-C_{m-mis}({\cal W}^{G,n},{\cal V}^{G,n})   \nonumber                       \\
&& \le\max \{ \limsup_{n \rightarrow \infty}  \frac{1}{n} \log \log |{\cal W}^{G,n}|,\limsup_{n \rightarrow \infty} \frac{1}{n}   \log \log |{\cal V}^{G,n}| \}.
\end{eqnarray}
The bound is the best possible, that is, there exists a GAVWC such that the bound is tight.
\end{theorem}
The proof is provided in Appendix \ref{sec: proof of theorem question 3}.
The requirement \eqref{eq_vn} of Theorem \ref{the: question 3} is a very weak condition. It is sufficient to satisfy it to have an arbitrarily large $d$ and a small constant $v<1$ such that $v_n \ge v^{n^d}$, whereas typically $v_n$ is exponentially decreasing as $n$ increases.

We have seen an example that the upper bound in (\ref{eq_thm2}) is tight. But in the next example, we shall see that it is not always tight. 
 
\begin{example}
     Let $(W,V)$ be an ordinary wiretap channel (i.e., singly discrete memoryless wiretap channel), whose single-letter capacity formula is known. Its capacities of average error and strongly secure codes and maximum error and semantically secure codes are the same (which actually is a consequence of Theorem \ref{thmavwc}, because a single wiretap channel is an ``AVWC with single state.") That is, $C_{a-str}( W,V)=C_{m-mis}(W, V)$
 
 Let ${\cal Z}$ be the output alphabet of $V$ and ${\cal Q}_n$ be the set of stochastic matrices (or channels) from ${\cal Z}^n$ to ${\cal Z}^n$ , which are not necessarily memoryless. For all $n$, let
 \[({\cal W}^{G,n}, {\cal V}^{G,n})=\{ (W^{n}, V^{n} \circ  Q_n): Q_n \in {\cal Q}_n\},  \mbox{ where  $V^{n} \circ Q_n({\bf z}|{\bf x})=\sum_{{\bf z}'}W^n({\bf z}'|{\bf x})Q_n({\bf z}|{\bf z}')$}.\]
 Then $|{\cal V}^{G,n}|=\infty$. Notice that the $n$ dimensional identity matrix ${\bf I}_n \in {\cal Q}_n$ and $ V^n \circ {\bf I}_n=V^n$. Therefore, $(W^n,V^n) \in ({\cal W}^{G,n}, {\cal V}^{G,n})$ and so $C_{a-str}({\cal W}^{G,n}, {\cal V}^{G,n})\le C_{a-str}( W,V)$ and  $C_{m-mis}({\cal W}^{G,n}, {\cal V}^{G,n})\le C_{m-mis}( W,V)$. On the other hand, by the data processing inequality, for all random variables  $U$ with all probability distributions, under arbitrary random encoding $I(U; {\bf Z}(V^n\circ Q_n)) \le I(U; {\bf Z}(V^n))$ for all $Q_n \in {\cal Q}_n$. Hence, an average error and strongly secure code for $(W,V)$ is an average error and strongly secure code for $({\cal W}^{G,n}, {\cal V}^{G,n})$  and a maximum error and semantically secure code for $(W,V)$ is a  maximum error and semantically secure code for $({\cal W}^{G,n}, {\cal V}^{G,n})$. Thus,
\[C_{a-str}({\cal W}^{G,n}, {\cal V}^{G,n})=C_{m-mis}({\cal W}^{G,n}, {\cal V}^{G,n}) =C_{a-str}( W,V)=C_{m-mis}(W, V) .\] 
\end{example}
The following lemma is an important auxiliary result to prove Theorem \ref{thmavwc}.
\begin{lemma} \label{lemma_semaim}
Let $(\Phi_n,\psi_n)$ be 
%a sequence codes 
an average error and strongly secure code for $({\cal W}^{G,n}, {\cal V}^{G,n}),$
%n=1,2,\ldots ,$ 
with message set ${\cal U}^{G,n}$, satisfying (\ref{eq_ex01}), (\ref{eq_ex02}), and (\ref{eq_ex02a}). Suppose that 
%for all $n$,
positive integers $J_n$ and $K_n$ and positive real numbers $A_n$ and $B_n$ satisfy the following conditions: $\frac{J_nK_n}{|{\cal U}^{G,n}|}:=\beta_n < \frac{1}{4}$,
\begin{equation}\label{eq_mlmm01}
K_n\mu \le \frac{A_n}{4} \mbox{ and } K_n\lambda \le \frac{B_n}{4}
\end{equation}
\begin{equation}\label{eq_mlmm02}
|{\cal V}^{G,n}|J_ne^{-\frac{\delta_n^{-1}A_n}{4}} < \frac{1}{4}, |{\cal W}^{G,n}| J_n e^{-\frac{B_n}{4}} < \frac{1}{4} \mbox{ and } J_ne^{-\frac{K_n \beta_n}{32}} < \frac{1}{4}.
\end{equation}
Then for all sufficiently large $n$, there exists a random encoding code $(\Upsilon_n,\nu_n)$ with a message set $\{1,2,\ldots, J_n\}$, a random encoder $\Upsilon_n$ and a deterministic decoder $\nu_n$ such that for all $W^{G,n} \in {\cal W}^{G,n}$ and all  $j \in \{1,2,\ldots, J_n\}$
\begin{equation} \label{eq_mlmm03}
W^{G,n}(\nu_n({\bf Y})\not=j|\Upsilon_n(j)) \le \frac{B_n}{K_n(1-\frac{3\beta_n}{2})}
\end{equation}
and
\begin{equation} \label{eq_mlmm04}
I(\tilde{U};\tilde{\bf Z}(V^{G,n}))\le \frac{A_n}{K_n(1-\frac{3\beta_n}{2})},
\end{equation}
for all $V^{G,n} \in {\cal V}^{G,n}$ and random message $\tilde{U}$ with any probability distribution on the message set $\{1,2,\ldots, J_n\}$, where $\tilde{\bf Z}(V^{G,n})$ is the random output of $V^{G,n}$ for message $\tilde{U}$ when the code $(\Upsilon_n,\nu_n)$ is applied.
\end{lemma}
The proof is given in Appendix \ref{sec: proof of the lemma}. The lemma bounds the decoding error and information leakage with arbitrary message distribution, using a random encoding code constructed based on a given average error and strongly secure code. The GAVWC considered in the lemma and the following theorem is general. The only requirement is that the input and output alphabets are exponentially increasing as $n$ grows, which is very natural, because without it, we could not even define the transmission rate and capacity.

\section{Concluding Remarks and Discussions}
This paper addresses three questions regarding secure communication over general, arbitrarily varying wiretap channels. We show that the strong secrecy capacity and semantic secrecy capacity of arbitrarily varying wiretap channels are always the same, and the equality still holds for general arbitrarily varying wiretap channels when the sizes of the communication systems are sub-double-exponential. When such a condition does not hold, we prove by constructing an example that the semantic secrecy capacity can be strictly smaller than the strong secrecy capacity. Furthermore, we characterize the gap between strong secrecy capacity and semantic secrecy capacity of GAVWC in terms of the sizes of the communication systems.

\subsection{Comparison of capacity results with different criteria}
When solving these questions, we are treating two separate problems, reliable transmission over the main channel and the security on the wiretap channel, simultaneously. We set the main channels in the two examples as single channels, because we are mainly concerned with the security on the wiretap channel. One may set the wiretap channel as a single completely noisy channel, or simply assume that there is no eavesdropper at all, if he is only interested in the error probability criterion of the main channels. Thus, the two theorems provide us with the bounds of the gap between the maximum error probability and the average error probability capacities of codes with a random encoder for AVC and ``GAVC". In particular, by Theorem \ref{thmavwc}, for the capacity of AVC for random encoding codes, the criteria of probability of error make no difference. But this observation is not new. It appeared  in \cite{ahlswede1978elimination} (also on p. 220 in \cite{csiszar2011information}). On the other hand, actually, in light of the idea in the previous section, we may construct an example of GAVC with $|{\cal W}^{G,n}| ={2^{n} \choose 2^{na}}$  and the difference of the capacity of maximum probability of error and the capacity of average probability of error for random encoding codes is $a$, as follows. That is, the bound for the main channel is tight. Namely, let 
 ${\cal W}^{G,n}:=\{W_{\Theta}: \Theta \in {\{0,1\}^n \choose 2^{na}}\}$  be a GAVC with  the input alphabet for codes of length $n$,
${\cal X}^{G,n}=\{0,1\}^n$ and corresponding output alphabet ${\cal Y}^{G,n}=\{0,1\}^n \cup \{E\}$, such that  for all $\Theta \in {\{0,1\}^n \choose 2^{na}}, {\bf x} \in \{0,1\}^n, {\bf y} \in \{0,1\}^n \cup \{E\}$,
 \[
 W_{\Theta} ({\bf y}|{\bf x})= \left\{ \begin{array}{ll} 1 & \mbox{if ${\bf y}={\bf x}$ and ${\bf x}\ \not\in \Theta$}   \\
 0 & \mbox{if ${\bf y}\not={\bf x}$ and ${\bf x}\not \in \Theta$}   \\
 0 & \mbox{if ${\bf y} \in \{0,1\}^n$ and ${\bf x} \in \Theta$}   \\
 1  & \mbox{if ${\bf y}=E$ and ${\bf x} \in \Theta.$} \end{array} \right.
 \]
It is easy to verify that for this GAVC, the capacities of  random encoding codes with average probability of error and with maximum probability of error are $1$ and $1-a$, respectively. We omit the details, because it is almost the same as the proof in the previous section.

 Throughout the proofs of the lemma and theorems, neither the statistics of the channels nor the construction of codes are involved. We just show that from a (given) codebook of an average error and a strongly secure code, one can extract a maximum error and a semantically secure code. Thus, by simply replacing classical information quantities with quantum, the analogous quantum results for quantum GAVWC can be done essentially without changing anything. By slightly adjusting the proofs according to the problem setting, extending the results to other secure problems (e.g., classical and quantum random secret key generation) should not be much more difficult either .

We consider the worst case of secure communication, the worst channel, and the worst distribution of a random message when we apply the semantically secure criterion to GAVWC. This is equivalent to allowing the eavesdropper to choose  not only the channel to attack but also the distribution of the attacked message, according to the coding scheme. The legal communicators have to find a code to protect the message against all these attacks.
On the other hand, the eavesdropper can only choose the channel, but not the distribution of the random message, if the strongly secure criterion is applied.
We have seen the power of the attack by an eavesdropper in the former case, in the example in the previous section. The example in this section gives us an opposite extreme case. There exists the best wiretap channel for eavesdroppers, or the worst channel for the communicators, ($ V^n \circ {\bf I}=V^n$). No matter how the random message is distributed, the eavesdropper has no better choice.  So the game is much easier for communicators, and they only need to consider the worst channel for them. But the real world is not always like  the example.  Intuitively, it seems to us that the eavesdropper would be at a very advantageous position if we use a semantically secure criterion. Surprisingly, our results show that the eavesdropper may not take any advantage in rate at all, unless the number of his choices double exponentially increases or faster as $n$ grows. Hence, we would not need to care whether the secure criterion is strong or semantic if the choices of the eavesdropper are sub-double-exponential. Another open question left is the positivity condition of the GAVC. We have briefly discussed that for the AVC, which has already been fully addressed in \cite{csiszar1988capacity}\cite{ericson1985exponential}. However, it is still not clear if a similar property of symmetrizability still holds for GAVCs.

\subsection{Insights into cryptographic problems}
\begin{figure}
    \centering
    \includegraphics[scale=0.6]{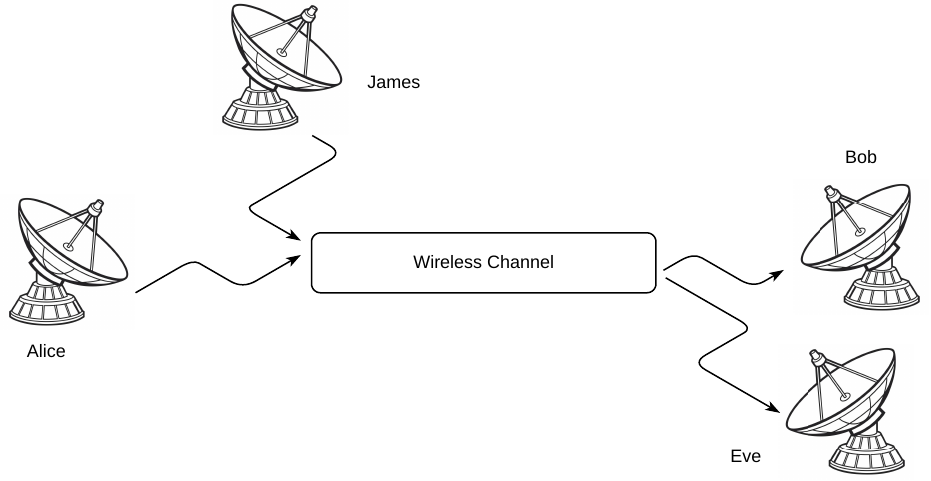}
    \caption{\footnotesize Communication system with a jammer}
    \label{fig:antenna}
\end{figure}
In physical communication channels, as shown in Fig. \ref{fig:antenna}, James can only select physical transmission signals as input to the channel. Since James is trying to disrupt the communication between Alice and Bob or to help Eve improve her security performance, he will choose a suitable physical jamming signal for this purpose. This choice naturally depends on James’s knowledge of the communication system. It is trivial—since we are considering a public, standardized communication system—that James knows the encoder (i.e., the encoding algorithm) used by Alice and the decoding algorithm used by Bob. However, he may also possess significantly more knowledge. Please discuss the diagram again now: the message itself or the message pair, etc. The assumption that James can directly manipulate Alice’s message—as is the case with standard cryptographic models such as tampering—does not generally hold true in physical communication systems. Instead, he can only influence the physical propagation conditions in a way that suits his purposes by using physical input signals appropriate to him, in order to prevent Bob from decoding the message. Ultimately, James’s choice of physical input signal influences the stochastic nature of the communication channel from Alice to Bob. This paper naturally assumes that James has complete knowledge of the channel from Alice to Bob—including his own impact on that channel—and the channel from Alice to Eve—including the impact of his own input. Thus, the information-theoretic security results are naturally not based on James and Eve being unaware of the channels. We also assume, of course, that Eve knows all the channels and the influence of James’s physical input signals on the channels. The analytic model in this work also applies to cases in which James or Eve can determine the transmitted message, which reflects different decoding and security metrics adopted in the criteria. However, the legitimate users, Alice and Bob, have no idea of what strategies the adversaries will take. This puts them in an unfavorable position, as for both decoding error and security metrics, they have to consider the worst case, although from a practical viewpoint, there is a trade-off between adversaries' impact on reliability and security.   It is proved in our results that when James has sub-double-exponential choices of jamming strategies, the maximal error and semantic/distinguishing/mutual information security capacities are the same as the average error and strong secrecy capacity. When his choices are double-exponential or more, the maximal error and semantic/distinguishing/mutual information security capacities are still the same, although they can be strictly smaller than the average error and strong secrecy capacity.  Hence, our results are useful  for the practical design of the coding schemes for secure communication over wiretap channels, which is the most fundamental problem of building a communication system, as coding for wiretap channels is similar to cryptographic primitives. 

Such research has already been conducted, as shown in \cite{voichtleitner2026experimental}, where the maximal likelihood attack strategy, known as the optimal attacking strategy, was examined on the additive Gaussian wiretap channels under the semantic security constraint. However, for real systems, the wiretap channel is only a small component; at the system level, much more complex attacks can occur. Real protocols routinely face active attackers who can submit chosen ciphertexts and learn from decryption behavior, errors, timing, or side channels. For that reason, IND-CPA is often insufficient in practice. Modern confidentiality targets typically require adaptive chosen-ciphertext security (IND-CCA2). These attacks target the implementation of the protocols, among other things, but they can also reveal new vulnerabilities in the modeling of the primitiveS. Accordingly, a new security measure called 'non-malleability' was proposed, and the connection between non-maleability and wiretap channels with these stronger types of adversary models is still an open research question.

\appendices

\section{proof of lemma \ref{lem: equivalence}}\label{sec: proof of lem equivalence}
We refer the condition $\max_{V^{G,n},P_U}\sum_u P_U(u)D_{V^{G,n}}(u) \to 0$ to the mutual information security (MIS) as suggested in \cite{bellare2012semantic} and denote it by $\mathrm{Adv}^{MIS}(\mathcal{V}^{G,n})$.
\subsection{DS $\to$ SS}
Let $\mathcal{A}^{SS}$ be the attacking strategy of the adversary targeting the semantic security with an arbitrary function $f$ on $\mathcal{U}$. The adversary constructs its attack on the distinguishing security $\mathcal{A}^{DS}$ as follows: For an arbitrary pair $(u_0,u_1)$ and $V^{G,n}\in\mathcal{V}^{G,n}$, it runs $\mathcal{A}^{SS}$ to the ciphertext $\textbf{Z}(V^{G,n})$. Note here we omit the second argument of $\textbf{Z}$ as the selected message is unknown to the adversary. The attacking $\mathcal{A}^{SS}$ outputs a value $v$ and the adversary sets
\begin{align}
    \mathcal{A}^{DS}(\textbf{Z}(V^{G,n})) = \left\{
        \begin{aligned}
            &1,\;\;v=f(u_1),\\
            &0,\;\;\text{otherwise}
        \end{aligned}
    \right.
\end{align}
Let $U,U_0,U_1$ be three independent but identically distributed random variables. It follows that
\begin{align}
    &Pr\{\mathcal{A}^{DS}(U_0,U_1,\textbf{Z}(V^{G,n},U_1))=1\} =Pr\{\mathcal{A}^{SS}(\textbf{Z}(V^{G,n},U)) =f(U)\},\\
    &Pr\{\mathcal{A}^{DS}(U_0,U_1,\textbf{Z}(V^{G,n},U_0))=1\} \leq \max_{\mathrm{S}} Pr\{\mathrm{S}(\mathcal{U}^{G,n}) = f(U)\}
\end{align}
by the fact that $U,U_0,U_1$ are independently and identically distributed. It follows that for every $V^{G,n} \in \mathcal{V}^{G,n}$,
\begin{align}
    &Pr\{\mathcal{A}^{SS}(\textbf{Z}(V^{G,n},U)) =f(U)\} - \max_{\mathrm{S}} Pr\{\mathrm{S}(\mathcal{U}^{G,n}) = f(U)\}\\
    &\leq Pr\{\mathcal{A}^{DS}(U_0,U_1,\textbf{Z}(V^{G,n},U_1))=1\} - Pr\{\mathcal{A}^{DS}(U_0,U_1,\textbf{Z}(V^{G,n},U_0))=1\}\\
    &=2Pr\{\mathcal{A}^{DS}(U_0,U_1,\textbf{Z}(V^{G,n},U_b))=b\} - 1\\
    &\leq \max_{u_0,u_1} 2Pr\{\mathcal{A}^{DS}(u_0,u_1,\textbf{Z}(V^{G,n},u_b))=b\} - 1
\end{align}
Taking the maximum over all adversary strategies and all distributions of $U$ completes the proof.

\subsection{SS $\to$ DS}
For any pair of messages $(u_0,u_1)$, define $U^{u_0,u_1}$ to be a random variable uniformly distributed on $\{u_0,u_1\}$. It follows that
\begin{align}
    \frac{1}{2}\mathrm{Adv}^{DS}(\mathcal{V}^{G,n}) &= \max_{u_0,u_1,V^{G,n}}\max_{\mathcal{A}^{DS}}\left(Pr\left\{\mathcal{A}^{DS}(u_0,u_1,\textbf{Z}(V^{G,n},u_b)) = b\right\} - \frac{1}{2}\right)\\
    &=\max_{u_0,u_1,V^{G,n}}\max_{\mathcal{A}^{SS}} Pr\{\mathcal{A}^{SS}(\textbf{Z}(V^{G,n},U^{u_0,u_1}))=U^{u_0,u_1}\} - \max_{\mathrm{S}} Pr\{\mathrm{S}(\mathcal{U}^{G,n}) = U^{u_0,u_1}\}\\
    &\leq \mathrm{Adv}^{SS}(\mathcal{V}^{G,n}),
\end{align}
where the second equality follows because when $U$ is distributed on a set with only two elements, and hence finding the index is equivalent to finding the element.

\subsection{MIS $\to$ DS}
By the definition of mutual information and Pinsker's inequality, 
\begin{align}
    \mathrm{Adv}^{MIS}(\mathcal{V}^{G,n}) &= \max_{P_U,V^{G,n}} I(U;\textbf{Z}(V^{G,n},U))\\
    &=\max_{P_U,V^{G,n}} D(P_{U\textbf{Z}(V^{G,n},U)}||P_UP_{\textbf{Z}(V^{G,n},U)})\\
    &\geq \max_{P_U,V^{G,n}} 2 \mathrm{SD}\left(P_{U\textbf{Z}(V^{G,n},U)},P_UP_{\textbf{Z}(V^{G,n},U)}\right)^2, 
\end{align}
where $D(\cdot || \cdot)$ is the KL divergence, $P_U$ and $P_{\textbf{Z}(V^{G,n},U)}$ is the marginal of the joint distribution $P_{U\textbf{Z}(V^{G,n},U)}$ defined by $P_U P_{\textbf{X}|U}V^{G,n}$ for a given $V^{G,n}.$ Furthermore, by \cite[Lemma 4.4]{bellare2012cryptographic}, for any given $V^{G,n}\in\mathcal{V}^{G,n}$ and a pair of messages $(u_0,u_1)$, let $U$ be a random variable uniformly distributed on $\{u_0,u_1\}.$ It follows that
\begin{align}
    \mathrm{SD}(P_{U\textbf{Z}(V^{G,n},U)}, P_UP_{\textbf{Z}(V^{G,n},U)}) = \frac{1}{2} \mathrm{SD}(g(u_0),g(u_1)),
\end{align}
where $g:\mathcal{U}^{G,n} \to \mathcal{Z}^{G,n}$ is a transform such that $\textbf{Z}(V^{G,n},M) = g(M)$. Let $(u_0^*,u_1^*)$ and $V^{G,n*}$ be the pair of message and realization of the channel such that $\mathrm{Adv}^{DS}(\mathcal{V}^{G,n})$ is achieved, and $U^{u_0^*,u_1^*}$ be a random variable uniformly distributed on $\{u_0^*,u_1^*\}$. It follows that
\begin{align}
    &\frac{1}{2} \mathrm{Adv}^{DS}(\mathcal{V}^{G,n})\\
    &=2\cdot \frac{1}{4} \cdot \mathrm{SD}(\textbf{Z}(V^{G,n*},u_0^*),\textbf{Z}(V^{G,n*},u_1^*))^2\\
    &=2\cdot \mathrm{SD}(P_{U^{u_0^*,u_1^*}\textbf{Z}(V^{G,n*},U^{u_0^*,u_1^*})}, P_UP_{\textbf{Z}(V^{G,n*},U^{u_0^*,u_1^*})})^2\\
    &\leq \mathrm{Adv}^{MIS}(\mathcal{V}^{G,n}),
\end{align}
where the last inequality follows because $U^{u_0^*,u_1^*}$ falls into the range of the maximisation of $\mathrm{Adv}^{MIS}$.

\subsection{DS $\to$ MIS}

By \cite[Lemma 4.8]{bellare2012cryptographic}, for any $V^{G,n}\in\mathcal{V}^{G,n}$,
\begin{align}\label{def: psd}
    \mathrm{PSD}(U,\textbf{Z}(V^{G,n},U)):= \max_{u_0,u_1} \mathrm{SD}(P_{\textbf{Z}(V^{G,n},U=u_0)},P_{\textbf{Z}(V^{G,n},U=u_1)})\geq \mathrm{SD}(P_{\textbf{Z}(V^{G,n},U)},P_{\textbf{Z}(V^{G,n},u)})
\end{align}
for any $u$, where $P_{\textbf{Z}(V^{G,n},U)} = \sum_{u}P_U(u)P_{\textbf{Z}(V^{G,n},u)}.$  We can write
\begin{align}
    I(U;\textbf{Z}(V^{G,n},U)) &= H(\textbf{Z}(V^{G,n},U)) - H(\textbf{Z}(V^{G,n},U)|U)\\
    &=\sum_u P_U(u) \left( H(\textbf{Z}(V^{G,n},U)) - H(\textbf{Z}(V^{G,n},u)|U=u^*) \right)\\
    &\leq H(\textbf{Z}(V^{G,n},U)) - H(\textbf{Z}(V^{G,n},u^*)|U=u^*),
\end{align}
where
\begin{align}
    u^* = \mathop{\arg\max}_{u} H(\textbf{Z}(V^{G,n},U)) - H(\textbf{Z}(V^{G,n},u)|U=u).
\end{align}
Let $\epsilon = \mathrm{Adv}^{DS}(\mathcal{V}^{G,n}).$ By \cite[Lemma 4.6]{bellare2012cryptographic} we have
\begin{align}
    &H(\textbf{Z}(V^{G,n},U)) - H(\textbf{Z}(V^{G,n},u^*)|U=u^*)\\
    &\leq 2\mathrm{SD}(P_{\textbf{Z}(V^{G,n},U)},P_{\textbf{Z}(V^{G,n},u^*)}) \cdot \log \frac{|\mathcal{Z}^{G,n}|}{\mathrm{SD}(P_{\textbf{Z}(V^{G,n},U)},P_{\textbf{Z}(V^{G,n},u^*)})}\\
    &\leq 2\mathrm{PSD}(U,\textbf{Z}(V^{G,n},U)) \cdot \log \frac{|\mathcal{Z}^{G,n}|}{\mathrm{PSD}(U,\textbf{Z}(V^{G,n},U))}\\
    &\leq 2\epsilon \cdot \log \frac{|\mathcal{Z}^{G,n}|}{\epsilon},
\end{align}
where the inequalities follow from \eqref{def: psd} and the fact that the function $f(x)=2x\log N/x$ is a monotonically increasing function of $x$. As the inequalities hold for all $V^{G,n}$ and $P_U$, we have
\begin{align}
    \mathrm{Adv}^{MSI}(\mathcal{V}^{G,n})\leq 2\epsilon \cdot \log \frac{|\mathcal{Z}^{G,n}|}{\epsilon},
\end{align}
where $\epsilon = \mathrm{Adv}^{DS}(\mathcal{V}^{G,n}).$
\section{proof of lemma \ref{lemma_semaim}}\label{sec: proof of the lemma}
Let  $U$ be a uniform random message on ${\cal U}^{G,n}$ and 
%${\bf Y}(W_n)$ and 
${\bf Z}(V^{G,n})$ be the random output of 
%$W_n$ and 
$V^{G,n}$ for $U$, 
%respectviely, 
when the code  $(\Phi_n,\psi_n)$ is applied.
%At first, we note that for an uniform random message $U$ on ${\cal U}_n$, code  $(\Phi_n,\psi_n)$, and output ${\bf Z}(V_n)$ of $V_n$ for $U$ and  $(\Phi_n,\psi_n)$
Then by (\ref{eq_ex01}), we have that
\begin{equation} \label{eq_e}
\begin{aligned}
    &\mathbb{E} W^{G,n}(\psi_n({\bf Y}) \not= U|\Phi_n(U))\\
    &=\sum_{u \in {\cal U}^{G,n}}\frac{1}{|{\cal U}^{G,n}|} W^{G,n}(\psi_n({\bf Y})\not=u|\Phi_n(u)) \le  \lambda
\end{aligned}
\end{equation}
by (\ref{eq_ex02}), we have 
 \begin{equation} \label{eq_i}
 \begin{aligned}
     &\mathbb{E} D_{V^{G,n}}(U)\\
     &=\sum_{u \in {\cal U}^{G,n}}\frac{1}{|{\cal U}^{G,n}|}\sum_{\bf z} V^{G,n}({\bf z}|\Phi_n(u))\log \frac{V^{G,n}({\bf z}|\Phi_n(u))}{\sum_{v}\frac{1}{|{\cal U}^{G,n}|}V^{G,n}({\bf z}|\Phi_n(v))}\\
     &=I(U;{\bf Z}(V^{G,n}))\le \mu.
 \end{aligned}
 \end{equation}

Let $U_{j,k}, j=1,2,\ldots,J_n, k=1,2,\ldots, K_n$ be randomly independently, and uniformly distributed on ${\cal U}^{G,n}$. Then by Chernov's bound, (\ref{eq_ex02a}), and (\ref{eq_i}), we have that for all $j \in \{1,2,\ldots,J_n\}, V^{G,n} \in {\cal V}^{G,n}$ 
\begin{eqnarray} \label{eq_ex03}
&& Pr \{\sum_{k=1}^{K_n} D_{V^{G,n}}(U_{j,k}) >A_n \}  \nonumber \\
&& \le e^{-\delta_n^{-1} A_n} \mathbb{E}e^{\delta_n^{-1} \sum_{k=1}^{K_n} D_{V^{G,n}}(U_{j,k}) }  \nonumber \\
&&= e^{-\delta_n^{-1} A_n}\prod_{k=1}^{K_n} \mathbb{E}e^{\delta_n^{-1}  D_{V^{G,n}}(U_{j,k}) }   \nonumber \\
&&\le e^{-\delta_n^{-1} A_n}\prod_{k=1}^{K_n}[1+  \mathbb{E} e \delta_n^{-1}  D_{V^{G,n}}(U_{j,k}) ]         \nonumber \\
&&= e^{-\delta_n^{-1} A_n}[1+e\delta_n^{-1}  \mathbb{E}    D_{V^{G,n}}(U) ]^{K_n}         \nonumber \\
&& \le e^{-\delta_n^{-1} A_n}(1+e \delta_n^{-1} \mu)^{K_n}    \nonumber \\
&&\le e^{-\delta_n^{-1}[A_n-eK_n\mu]} <e^{-\frac{\delta_n^{-1}A_n}{4}},
\end{eqnarray}
if $K\mu \le \frac{A_n}{4}$, where the first inequality is Chernov bound; the first equality holds because  $U_{j,k}, j=1,2,\ldots,J_n, k=1,2,\ldots, K_n$ are independent; the second inequality holds because by (\ref{eq_ex02a}) $ \delta_n^{-1}  D_{V^{G,n}}(U_{j,k}) \le 1 \ (a.s.)$ and $e^z \le 1+ez$ for all $z \in [0,1]$; and by  (\ref{eq_i}), the third inequality holds; and the fourth inequality follows from the inequality $1+z \le e^z$. 

Similarly, as $W^{G,n}(\psi({\bf Y})\not=u|\Phi(u))$ 
%$W_n(\psi({{\bf Y}(W_n)})\not=u|\Phi(u))$
 is upper bounded by one, for all $u$, by Chernov bound and (\ref{eq_e}), we have that for all $j \in \{1,2,\ldots,J\}, V^{G,n} \in {\cal V}^{G,n}$
\begin{eqnarray}\label{eq_ex04}
&&Pr\{\sum_{k=1}^{K_n} W^{G,n}(\psi({\bf Y}) )\not= U_{j,k} |\Phi(U_{j,k}) )> B_n\} \nonumber \\
&& \le e^{-B_n} \mathbb{E}e^{\sum_{k=1}^{K_n} W^{G,n}(\psi({\bf Y}) \not= U_{j,k} |\Phi(U_{j,k}) )}                                       \nonumber \\     
&&=  e^{-B_n} \prod_{k=1}^{K_n} \mathbb{E}e^{ W^{G,n}(\psi({\bf Y}) \not= U_{j,k} |\Phi(U_{j,k}) )} \nonumber \\       
&&\le e^{- B_n}\prod_{k=1}^{K_n}[1+  \mathbb{E} e W^{G,n}(\psi({\bf Y}) \not= U_{j,k} |\Phi(U_{j,k}) ) ]         \nonumber \\
&&= e^{-B_n}[1+e  \mathbb{E}  W^{G,n}(\psi({\bf Y}) \not= U |\Phi(U) ) ]^{K_n}         \nonumber \\
&& \le e^{-B_n}(1+e \lambda)^{K_n}    \nonumber \\                        
&&< e^{-(B_n -eK_n\lambda)} \le e^{-\frac{B_n}{4}},
\end{eqnarray}
if $K_n\lambda \le \frac{B_n}{4}$.
%where ${\bf Y}(W_n)$ is the random output of $W_n$ for the uniform $U_{j,k}$, when the code  $(\Phi_n,\psi_n)$ is applied.

Next, for any $\bar{\cal U}\subset{\cal U}^{G,n}$ with $\frac{|\bar{\cal U}|}{|{\cal U}^{G,n}|} \le \beta_n$, and $j=1,2 \ldots, J_n$, we let
\[F_k=\left\{\begin{array}{ll} 
1 & \mbox{if $U_{j,k} \in \bar{\cal U}$}  \\
0 & \mbox{else,}
\end{array}  \right.\]
and apply
\begin{lemma} \label{lemma_chernoff} %({\it Chernoff Bound}) 
( Lemma 3 in \cite{cai2013localized} and also Lemma 2 in \cite{boche2018message})
Let
$F_1,F_2, \dots, F_L$ be i.i.d. random binary sequences taking values in $\{0,1\}$, with
$Pr(F_l=1)=p$. Then for all $\alpha \in (0,1), p\le p_1$
\begin{equation}\label{eq_lemmach1}
Pr\{\sum_{l=1}^LF_l > Lp_1(1+\alpha)\} <e^{-\frac{{\alpha}^2}{8}Lp_1}.         \nonumber
\end{equation}
%and
%\begin{equation}\label{eq_lemmach2}
%Pr\{\sum_{l=1}^LB_l < Lp_0(1-\alpha)\}
%<e^{-\frac{{3\alpha}^2}{8}Lp_0}.
%\end{equation}
\end{lemma}
for $L=K_n$, 
%$F_k=1$, 
%if $U_{j,k} \in \bar{\cal U}$ and otherwise $F_k=0$, 
$p_1=\beta_n$ 
and $\alpha=\frac{1}{2}$ and obtain
\begin{eqnarray} \label{eq_ex05}
&&Pr\{|\{k: U_{j,k} \in\bar {\cal U}, k \in \{1,2,\ldots, K_n\}\}| >\frac{3K_n \beta_n}{2}\}  \nonumber  \\
&&=Pr\{\sum_{l=1}^LF_l > K_n \beta_n(1+\frac{1}{2})\}< e^{-\frac{K_n \beta_n}{32}},
\end{eqnarray}
for  $j=1,2 \ldots, J_n$.

Now we let us recall the condition (\ref{eq_mlmm02})
%\[|{\cal V}_n|Je^{-\frac{\delta_n^{-1}A_n}{4}} < \frac{1}{4}, |{\cal V}_n| J e^{-\frac{B_n}{4}} < \frac{1}{4}, \mbox{ and },Je^{-\frac{K \beta_n}{32}} < \frac{1}{4}\]
for $\frac{J_nK_n}{|{\cal U}^{G,n}|}=\beta_n$. Then we begin with $j=1$ and find realizations $u_{j,1}, u_{j,2}, \ldots,u_{j,K}$ of $U_{j,1}, U_{j,2}, \ldots,U_{j,K}$,  step by step 
%and apply (\ref{eq_ex03}),(\ref{eq_ex04}), and (\ref{eq_ex05})
%(\ref{eq_ex03}),(\ref{eq_ex02}), and (\ref{eq_ex03}) 
recursively for $j=1,2,\ldots, J_n$,
%untill $j=J_n$,with a probability at least $\frac{1}{4}$, 
such that for all $j$ and all $V^{G,n} \in {\cal V}^{G,n}$,
\begin{equation} \label{eq_ex06}
\sum_{k=1}^{K_n} D_{V^{G,n}}(u_{j,k}) \le A_n
\end{equation}
\begin{equation} \label{eq_ex07}
\sum_{k=1}^{K_n} W^{G,n}(\psi({\bf Y})\not= u_{j,k} |\Phi(u_{j,k}) ) \le B_n
\end{equation}
and
\begin{equation} \label{eq_ex08}
|\{k: u_{j,k} \in\bar {\cal U}_{j-1}, k \in \{1,2,\ldots, K_n\}\}| \le \frac{3K_n \beta_n}{2} 
\end{equation}
for $\bar{\cal U}_{j-1}:=\{u_{j',k'}, j'=1,2,\ldots, j-1, k'=1,2,\ldots,K_n\}$ for $j>1$ , and  $\bar{\cal U}_0=\emptyset$.
%with a positive probability, where $\bar{\bf Y}_{j,k}(W_n)$ is the random output of $W_n$ when the realization $u_{j,k}$ of $U_{j,k}$ works as a message for $W_n$.
%when $j=1$, $\bar{\cal U}_{j-1}$ is understood ad the empty set.

We say that the  step $j$ is  successful with a positive probability if with a positive probability we may find a realization $u_{j,1}, u_{j,2}, \ldots,u_{j,K}$ of $U_{j,1}, U_{j,2}, \ldots,U_{j,K}$ satisfying (\ref{eq_ex06}), (\ref{eq_ex07}) and (\ref{eq_ex08}) at the step $j$.

To show that the procedure is feasible, we have to show that the first $j$ steps are successful with a positive probability, for $j=1,2,\ldots, J_n$, 
%there exist candidates of realization $u_{j,1}, u_{j,2}, \ldots,u_{j,K}$ of $U_{j,1}, U_{j,2}, \ldots,U_{j,K}$ satisfying (\ref{eq_ex06}), (\ref{eq_ex07}) and (\ref{eq_ex08})
%with a positive probability at each step $j$,
%We have to show that the successful probability to complete is at least $\frac{1}{4}$ recursively, 
by (\ref{eq_ex03}),(\ref{eq_ex04}), and (\ref{eq_ex05}) recursively, by induction.

 Indeed,
 % by (\ref{eq_ex03}),(\ref{eq_ex04}), and (\ref{eq_ex05}), 
  at the first step, by (\ref{eq_ex03}) and (\ref{eq_ex04}),  the probabilities to fail  for having a realization satisfying (\ref{eq_ex06}) for some $V^{G,n} \in {\cal V}^{G,n}$ and $j=1$, and to fail  for having a realization satisfying (\ref{eq_ex07}) for some $W^{G,n} \in {\cal W}^{G,n}$ and $j=1$, %and  for to fail  for having a realization satisfying (\ref{eq_ex08}) for $j=1$ 
  are at most $|{\cal V}^{G,n}|e^{-\frac{\delta_n^{-1}A_n}{4}}$, and $|{\cal W}^{G,n}| e^{-\frac{B_n}{4}}$, 
  %and $ e^{-\frac{K_n \beta_n}{32}}$ 
  respectively. 
  Due to $\bar{\cal U}_0=\emptyset$, (\ref{eq_ex08}) holds with probability one.
  Hence, the successful probability at the first step is at least 
 \begin{eqnarray}
 && 1-[|{\cal V}^{G,n}|e^{-\frac{\delta_n^{-1}A_n}{4}}+|{\cal W}^{G,n}|e^{-\frac{B_n}{4}}]    \nonumber \\
&&> 1-[|{\cal V}^{G,n}|e^{-\frac{\delta_n^{-1}A_n}{4}}+|{\cal W}^{G,n}|e^{-\frac{B_n}{4}}+e^{-\frac{K_n \beta_n}{32}}]. \nonumber 
 \end{eqnarray}
Having shown the successful probability at the first the $j-1$st steps is at least
 \[1-(j-1)[|{\cal V}^{G,n}|e^{-\frac{\delta_n^{-1}A_n}{4}}+|{\cal W}^{G,n}|e^{-\frac{B_n}{4}}+e^{-\frac{K_n \beta_n}{32}}],\]
we observe that at this situation, by (\ref{eq_ex03}),(\ref{eq_ex04}), and (\ref{eq_ex05}), at the $j$th step, the probabilities to fail  for having a realization satisfying (\ref{eq_ex06}) for some $V^{G,n} \in {\cal V}^{G,n}$ and $j$, to fail  for having a realization satisfying (\ref{eq_ex07}) for some $W^{G,n} \in {\cal W}^{G,n}$ and $j$, and  for to fail  for having a realization satisfying (\ref{eq_ex08}) for $j$ are at most $|{\cal V}^{G,n}|e^{-\frac{\delta_n^{-1}A_n}{4}}$,$|{\cal W}^{G,n}| e^{-\frac{B_n}{4}}$, and $ e^{-\frac{K_n \beta_n}{32}}$ respectively. Thus, by independence of $U_{j,k}$'s, the  successful probability to complete the first $j$ steps is at least
\begin{eqnarray}
&&\{1-[|{\cal V}^{G,n}|e^{-\frac{\delta_n^{-1}A_n}{4}}+|{\cal W}^{G,n}|e^{-\frac{B_n}{4}}+e^{-\frac{K_n \beta_n}{32}}]\} \times \nonumber \\
&&\times \{1-(j-1)[|{\cal V}^{G,n}|e^{-\frac{\delta_n^{-1}A_n}{4}}+|{\cal W}^{G,n}|e^{-\frac{B_n}{4}}+e^{-\frac{K_n \beta_n}{32}}]\}  \nonumber \\
&& \ge 1-j[|{\cal V}^{G,n}|e^{-\frac{\delta_n^{-1}A_n}{4}}+|{\cal W}^{G,n}|e^{-\frac{B_n}{4}}+e^{-\frac{K_n \beta_n}{32}}].\nonumber
\end{eqnarray}
We continue the procedure until $j=J_n$, and then show that the probability of completing the whole procedure successfully is at least
\[ 1-J_n[|{\cal V}^{G,n}|e^{-\frac{\delta_n^{-1}A_n}{4}}+|{\cal W}^{G,n}|e^{-\frac{B_n}{4}}+e^{-\frac{K_n \beta_n}{32}}]\]
which is at least $ \frac{1}{4}$ by the condition (\ref{eq_mlmm02}).

Let $\tilde{\cal U}_j= \{u_{j,k},k=1,2,\ldots, K_n\}\setminus \bar{\cal U}_{j-1}$, 
%$\tilde{\cal U}_j={\cal U}_j \setminus \bar{\cal U}_{j-1}$,
and then $\tilde{\cal U}_j \subset \{u_{j,k},k=1,2,\ldots, K_n\}$ and  by (\ref{eq_ex08}), 
\begin{equation}\label{eq_ex08a}
K_n(1-\frac{3\beta_n}{2}) \le |\tilde{\cal U}_j |\le K_n.
\end{equation}
Now, we take $\{1,2,\ldots,J_n\}$ as message set and define a new random encoder $\Upsilon_n$ and deterministic decoder $\nu_n$, such that for sending message $j \in \{1,2,\ldots, J_n\}$ the encoder randomly uniformly chooses a $ u \in \tilde{\cal U}_j$ and then send it by the older encoder $\Phi_n$. The decoder outputs $\hat{j}$, if $\psi({\bf z})\in \tilde{\cal U}_{\hat{j}}$, after it receives ${\bf z}$ from the main channel. As  the sets $\tilde{\cal U}_j$'s are pairwise disjoint, the decoder is well defined.
Thus, for all  $j \in \{1,2,\ldots, J_n\}$ and all $V^{G,n} \in {\cal V}^{G,n}$, by (\ref{eq_ex06}), (\ref{eq_ex08a}), and convexity of the divergence,
\begin{align}
    & \sum_{\bf z} V^{G,n}({\bf z}|\Upsilon_n(j))\log \frac{V^{G,n}({\bf z}|\Upsilon_n (j))}{\sum_{v}\frac{1}{|{\cal U}^{G,n}|}V^{G,n}({\bf z}|\Phi_n(v))}     \notag                       \\  
&=\sum_{\bf z} [\sum_{u \in \tilde{\cal U}_j}\frac{1}{|\tilde{\cal U}_j|}V^{G,n}({\bf z}|\Phi_n(u))]\log \frac{ [\sum_{u \in \tilde{\cal U}_j}\frac{1}{|\tilde{\cal U}_j|}V^{G,n}({\bf z}|\Phi_n(u))]}{\sum_{v}\frac{1}{|{\cal U}^{G,n}|}V^{G,n}({\bf z}|\Phi_n(v))}     \notag                       \\  
&\le \sum_{u \in \tilde{\cal U}_j}\frac{1}{|\tilde{\cal U}_j|}[\sum_{\bf z} V^{G,n}({\bf z}|\Phi_n(u)) \log \frac{ V^{G,n}({\bf z}|\Phi_n(u))}{\sum_{v}\frac{1}{|{\cal U}^{G,n}|}V^{G,n}({\bf z}|\Phi_n(v))}]     \notag                       \\  
&=\sum_{u \in \tilde{\cal U}_j}\frac{D_{V^{G,n}}(u)}{|\tilde{\cal U}_j|} \le \sum_{k=1}^{K_n}\frac{D_{V^{G,n}}(u_{j,k})}{|\tilde{\cal U}_j|}  \notag                       \\ 
\label{eq_ex09}&\le \frac{A_n}{|\tilde{\cal U}_j|}\le \frac{A_n}{K(1-\frac{3\beta_n}{2})},
\end{align}
where the first equality holds by definition of the code; the first inequality follows from the  convexity of the divergence; the second equality follows from (\ref{eq_D}); the second inequality holds, because $\tilde{\cal U}_j \subset \{u_{j,1},u_{j,2} \ldots u_{j,K_n}\}$ and $D_{V^{G,n}}(u) \ge 0$ for all $u$; by ({\ref{eq_ex06}) the third inequality holds; and finally the last inequality follows from (\ref{eq_ex08a}).
(\ref{eq_ex09}) with Csisz\'{a}r's Information Radius Lemma, together yields (\ref{eq_mlmm04}) for $\tilde{U}$ with all distribution.

Similarly,  by (\ref{eq_ex07}), (\ref{eq_ex08a}), for all  $j \in \{1,2,\ldots, J\}$ and all $W^{G,n} \in {\cal W}^{G,n}$,
\begin{align} 
&W^{G,n}(\nu_n({\bf Y})\not=j|\Upsilon_n(j)) = \sum_{u \in \tilde{\cal U}_j} \frac{1}{|\tilde{\cal U}_j|}W^{G,n}(\psi_n ({\bf Y}) \not= u|\Phi_n(u))\notag\\
& \le \sum_{k=1}^{K_n}\frac{1}{|\tilde{\cal U}_j|} W^{G,n}(\psi_n({\bf Y}) \not= u_{j,k} |\Phi_n(u_{j,k}) )\notag \\
\label{eq_ex10}&\le \frac{B_n}{|\tilde{\cal U}_j|}\le \frac{B_n}{K_n(1-\frac{3\beta_n}{2})}.
\end{align}
That is (\ref{eq_mlmm03}). Thus, our proof is completed.

\section{proof of theorem \ref{thmavwc}}\label{sec: proof of the thm}
\subsection{Proof of Part 1:} For an AVWC $({\cal W}, {\cal V})$ with the set of states ${\cal S}$, let $v_0:=\min_{x,s,z} \{ V(z|x,s) >0\}$.%  
Then $ |{\cal  W}^{G,n}|= |{\cal  V}^{G,n}|=|{\cal S}|^n$ and $v_n=v_0^n$. Thus, (\ref{eq_thmAVC}) holds and therefore (ii) yields (i).

\subsection{Proof of Part 2:}
By (\ref{eq_D}), for all $n, V^{G,n},u$ and all random encoder $\Phi$,
\begin{equation} \label{eq_dv}
\begin{aligned}
    &D_{V^{G,n}}(u)=\sum_{\bf z} V^{G,n}({\bf z}|\Phi_n(u))\log \frac{V^{G,n}({\bf z}|\Phi_n(u))}{\sum_{v}\frac{1}{|{\cal U}^{G,n}|}V^{G,n}({\bf z}|\Phi_n(v))}\\
    &\le \sum_{\bf z} V^{G,n}({\bf z}|\Phi_n(u))\log \frac{1}{v_n} = \log v_n^{-1},
\end{aligned}
\end{equation}
that is, (\ref{eq_ex02a}), for $\delta_n=-n\log v_0$. Hence, we can apply Lemma \ref{lemma_semaim} to $({\cal W}^{G,n}, {\cal V}^{G,n})$ by letting $\delta_n= \log v_n^{-1}$.

For all $\epsilon, \lambda, \mu >0$, there exists an average error and strongly secure code $(\Phi_n, \psi_n)$ with message set ${\cal U}^{G,n}$ and rate $\frac{1}{n} log {|{\cal U}^{G,n}|} >C_{a-str}(({\cal W},{\cal V}))-\frac{\epsilon}{3}$ such that for a uniform random message $U$ on ${\cal U}^{G,n}$, the average probability of error for all $W^{G,n} \in {\cal W}^{G,n}$, 
\begin{equation} \label{eq_ex01avc}
\sum_{u \in {\cal U}^{G,n}}\frac{1}{|{\cal U}^{G,n}|} W^{G,n}(\psi_n({\bf Y})\not=u|\Phi_n(u)) \le \frac{\lambda}{8},
\end{equation}
and the mutual information for all $V^{G,n} \in {\cal V}^{G,n}$, leaking to the eavesdropper
\begin{equation} \label{eq_ex02avc}
I(U;{\bf Z}(V^{G,n}))\le \frac{\mu}{8}.
%I(U;{\bf Z}(V_n))= \sum_{u \in {\cal U}_n}  D_{V_n}(u) \log \frac{D_{V_n}(u)}{\sum_{v \in {\cal U}_n}D_{V_n}(v)   } \le \mu_n   \rightarrow 0
\end{equation}
%Let $(\Phi_n,\psi_n),n=1,2, \ldots,$ be an average error and strongly secure codes for $({\cal W}, {\cal V})$ satisfying (\ref{eq_ex01}) and (\ref{eq_ex02}).

To apply Lemma \ref{lemma_semaim}, we let $K_n=2^{\frac{n\epsilon}{3}}, \beta_n=\frac{1}{n}$, and hence
\[\log J_n=\log|{\cal U}^{G,n}|-\frac{n\epsilon}{3} -\log n>n(C_{a-str}(({\cal W}^{G,n},{\cal V}^{G,n})-\epsilon),\]
if $n$ is sufficiently large. Let $A_n=\frac{1}{2}K_n \mu$ and $B_n=\frac{1}{2}K_n\lambda$, so that (\ref{eq_mlmm01}) is satisfied, after replacing $\mu$ and $\lambda$ by $\frac{\mu}{8} $ and $ \frac{\lambda}{8}$ respectively, in (\ref{eq_mlmm01}). Notice that by (\ref{eq_ex01avc}) and  (\ref{eq_ex02avc}), the roles of $\mu$ and $\lambda$ in the lemma have been taken by $\frac{\mu}{8} $ and $ \frac{\lambda}{8}$ respectively, and so now we only need to check (\ref{eq_mlmm02}) for the conditions of  Lemma \ref{lemma_semaim} for sufficiently large $n$. 
%Let the set of states of AVWC be ${\cal S}$, i.e., $|{\cal W}_n|=|{\cal V}_n |=|{\cal S}|^n$. 
Indeed, we have that
\begin{align*}
    &|{\cal V}^{G,n}|J_ne^{-\frac{\delta_n^{-1}A_n}{4}} =|{\cal V}^{G,n}|J_ne^{-\frac{\delta_n^{-1}K_n\mu}{8}} \\
    &\le 2^{-[\frac{2^{\frac{n\epsilon}{3}}\mu}{8\log v_n^{-1}}\log e-\log |{\cal U}^{G,n}|-\log |{\cal V}^{G,n}|]}< \frac{1}{4},
\end{align*} 
since  $J_n \le |{\cal U}^{G,n}|, \frac{2^{\frac{n\epsilon}{3}}}{\log v_n^{-1}}=2^{n(\frac{\epsilon}{3}-\frac{1}{n} \log \log v_n^{-1})}=2^{n(\frac{\epsilon}{3}+o(1))}, \log \log |{\cal V}^{G,n}|=o(n)$ by the assumption (\ref{eq_thmAVC}), and $\log |{\cal U}^{G,n}|$ is linearly increasing as $n$ is increasing. Similarly, by the assumption (\ref{eq_thmAVC})
\begin{align*}
    &|{\cal W}^{G,n}| J_n e^{-\frac{B_n}{4}}  =  |{\cal W}^{G,n}|J_ne^{-\frac{K_n\lambda}{8}}  \\
    &\le 2^{-[\frac{2^{\frac{n\epsilon}{3}}\lambda}{8}\log e-\log |{\cal U}^{G,n}|-\log |{\cal W}^{G,n}|]}< \frac{1}{4},
\end{align*}
and
\[J_ne^{-\frac{K_n \beta_n}{32}} \le 2^{-[\frac{2^{\frac{n\epsilon}{3}} }{32n}\log e-\log |{\cal U}^{G,n}|]} < \frac{1}{4},\]
if $n$ is sufficiently large. Thus,  Lemma \ref{lemma_semaim} is applied. Consequently, there exists a random encoding code $(\Upsilon_n, \nu_n)$ with a rate larger than $C_{a-str}(({\cal W}^{G,n},{\cal V}^{G,n})-\epsilon$ such that  (\ref{eq_mlmm03}) and (\ref{eq_mlmm04}) hold. 
Indeed, it is a  maximum error and semantically secure code, because by substitution of $A_n=\frac{1}{2}K_n \mu$ and $B_n=\frac{1}{2}K_n\lambda$ in (\ref{eq_mlmm03}) and (\ref{eq_mlmm04}), we obtain that
 \[W^{G,n}(\nu_n({\bf Y})\not=j|\Upsilon_n(j)) \le \frac{\lambda }{2(1-\frac{3\beta_n}{2})}< \lambda,\]
and 
\[I(\tilde{U};\tilde{\bf Z}(V^{G,n}))\le \frac{\mu}{2(1-\frac{3\beta_n}{2})}< \mu,\]
for sufficiently large $n$. That is,  $C_{a-str}({\cal W}^{G,n},{\cal V}^{G,n})$ is achievable by  maximum error and semantically secure codes, or $C_{a-str}({\cal W}^{G,n},{\cal V}^{G,n})\le C_{m-mis}({\cal W}^{G,n},{\cal V}^{G,n})$, which yields the theorem, because due to the fact that  a maximum error and semantically secure code is an average error and strongly secure code, the opposite inequality is trivial. 

\section{proof of theorem \ref{the: question 2}}\label{sec: proof of theorem question 2}
Let us consider the following communication system, $(W^{G,n}, {\cal V}^{G,n})_{n=1}^n$, where for each integer $n$, the main channel
 $W^{G,n}$ is a binary noiseless channel, from $n$-dimensional binary space $\{0,1\}^n$ to itself and the wiretap channel ${\cal V}^{G,n}$ is a set of channels labeled by $f(n)$-subsets of $\{0,1\}^n$ i.e., ${\cal V}^{G,n}:=\{V_{\Theta}: \Theta \in {\{0,1\}^n \choose f(n)}\}$ for an integer-valued function $f(n)$ with $\lim_{n\rightarrow \infty} \frac{nf(n)}{2^n} =0$, such that  for all $\Theta \in {\{0,1\}^n \choose f(n)}, {\bf x},{\bf y} \in \{0,1\}^n$,
 \begin{equation} \label{eq_01}
 V_{\Theta} ({\bf y}|{\bf x})= \left\{ \begin{array}{ll} 1 & \mbox{if ${\bf y}={\bf x}$ and ${\bf x} \in \Theta$}   \\
 0 & \mbox{if ${\bf y}\not={\bf x}$ and ${\bf x} \in \Theta$}   \\
 \frac{1}{2^n}  & \mbox{else.} \end{array} \right.
 \end{equation}
That is, for an input codeword in $\Theta$, $V_{\Theta}$ is the identity channel, and for inputs not in the subset $\Theta$, $V_{\Theta}$ is a completely noisy channel. A sender uses a random (or deterministic as a special case) encoding function $\Phi$ from a message set ${\cal M}$ to the input alphabet ${\cal X}^n=\{0,1\}^n$  to encode. The decoder is deterministic i.e., the decoding sets are a partition of output space $\{0,1\}^n$.
An eavesdropper knowing the coding scheme is allowed to freely choose a channel in ${\cal V}^{G,n}$ and access it. A decodable code with vanishing probability of error in the main channel, is strongly secure if for an uniformly distributed random message $M$, $\lim_{n \rightarrow \infty} \max_{\Theta} I(M;{\bf Y}(\Theta))=0$, where ${\bf Y}(\Theta)$ is random output observed by the eavesdropper by accessing the channel $V_{\Theta}$. It is semantically secure, if  $\lim_{n \rightarrow \infty} \max_{\Theta} I(M;{\bf Y}(\Theta))=0$ for any probability distribution of $M$. Moreover, we say that a decodable code for the main channel  is weakly semantically secure, if  $\lim_{n \rightarrow \infty} \max_{\Theta} \frac{1}{n} I(M;{\bf Y}(\Theta))=0$ for $M$ with any probability distribution. Obviously, a semantically secure code is weakly semantically secure.

At first, we consider a naive code with a rate of $1$ as a simple example.
%the strong security and so assume the random message $M$ is uniformly distributed on ${\cal M}$. 
That is, we let ${\cal M}=\{0,1\}^n$ and use the identity encoder i.e., $\Phi({\bf b})={\bf b}, {\bf b} \in \{0,1\}^n$ to encode. The decoder outputs ${\bf b}$ if it receives ${\bf b}$ from the main channel. Obviously, for the main channel, it is an error-free code.

Let us first consider its strong security and so assume the random message $M$ is uniformly distributed on ${\cal M}$. In this case, we can assume the message is a uniformly chosen codeword from the input alphabet $\{0,1\}^n$ and do not distinguish the random message $M$ and input codeword ${\bf X}$. 
%Since the main channel is noiseless, the idnetify mapping is an error free code for it. 
It is also easy to verify that it is strongly secure. In fact, by (\ref{eq_01}),  for all $\Theta \in {\{0,1\}^n \choose f(n)}, {\bf y} \in \Theta$, the output probability
\begin{eqnarray}
&&Pr\{{\bf Y}(\Theta)={\bf y}\}=\sum_{{\bf x}}Pr({\bf X}={\bf x}) V_{\Theta}({\bf y}|{\bf x})  \nonumber \\
&&=\sum_{{\bf x} \not\in \Theta}Pr({\bf X}={\bf x}) V_{\Theta}({\bf y}|{\bf x}) +\sum_{{\bf x} \in \Theta}Pr({\bf X}={\bf x}) V_{\Theta}({\bf y}|{\bf x}) 
\nonumber \\ 
&&=\sum_{{\bf x} \not\in \Theta}Pr({\bf X}={\bf x}) V_{\Theta}({\bf y}|{\bf x})+Pr({\bf X}={\bf y})V_{\Theta}({\bf y}|{\bf y})   \nonumber \\
&&= \frac{2^n-f(n)}{2^n} \frac{1}{2^n}+ \frac{1}{2^n}=\frac{1}{2^n}(2-\frac{f(n)}{2^n}) \nonumber
\end{eqnarray}
For ${\bf y} \not\in \Theta$, as $V_{\Theta}({\bf y}|{\bf x})=0$ for ${\bf x} \in \Theta$, we have
\begin{equation}
Pr\{{\bf Y}(\Theta)={\bf y}\}=\sum_{{\bf x} \not\in \Theta}Pr({\bf X}={\bf x}) V_{\Theta}({\bf y}|{\bf x})= \frac{2^n-f(n)}{2^n} \frac{1}{2^n}. \nonumber  \nonumber
\end{equation}
Hence, for all $\Theta$,
\begin{eqnarray}
&&I(M;{\bf Y}(\Theta))=\sum_{{\bf x}}\sum_{\bf y} Pr({\bf X}={\bf x}) V_{\Theta}({\bf y}|{\bf x})  \log \frac{ V_{\Theta}({\bf y}|{\bf x}) }{Pr(Y(\Theta)={\bf y})} \nonumber   \\
&&=\sum_{{\bf x} \not\in \Theta}\sum_{\bf y} Pr({\bf X}={\bf x}) V_{\Theta}({\bf y}|{\bf x})  \log \frac{ V_{\Theta}({\bf y}|{\bf x}) }{Pr(Y(\Theta)={\bf y})}   \nonumber    \\
&&+   \sum_{{\bf x} \in \Theta}\sum_{\bf y}Pr({\bf X}={\bf x}) V_{\Theta}({\bf y}|{\bf x})  \log \frac{ V_{\Theta}({\bf y}|{\bf x}) }{Pr(Y(\Theta)={\bf y})}                     \nonumber    \\    
&&=\sum_{{\bf x} \not\in \Theta}\sum_{{\bf y} \not\in \Theta} Pr({\bf X}={\bf x}) V_{\Theta}({\bf y}|{\bf x})  \log \frac{ V_{\Theta}({\bf y}|{\bf x}) }{Pr(Y(\Theta)={\bf y})}   \nonumber    \\
&&+\sum_{{\bf x} \not\in \Theta}\sum_{{\bf y} \in \Theta}Pr({\bf X}={\bf x}) V_{\Theta}({\bf y}|{\bf x})  \log \frac{ V_{\Theta}({\bf y}|{\bf x}) }{Pr(Y(\Theta)={\bf y})}   \nonumber    \\               
&&+ \frac{f(n)}{2^n} \log \frac{1}{\frac{1}{2^n}(2-\frac{f(n)}{2^n}) }  \nonumber   \\
&& =\frac{2^n-f(n)}{2^n} \frac{2^n-f(n)}{2^n} \log \frac{\frac{1}{2^n}}{ \frac{2^n-f(n)}{2^n} \frac{1}{2^n}}+ \frac{2^n-f(n)}{2^n} \frac{f(n)}{2^n} \log \frac{\frac{1}{2^n}}{\frac{1}{2^n}(2-\frac{f(n)}{2^n})}             \nonumber  \\
&&+ \frac{f(n)}{2^n} [n-\log(2- \frac{f(n)}{2^n})]      \nonumber  \\
%&& =\frac{2^n-f(n)}{2^n} \frac{1}{2^n} \log \frac{\frac{1}{2^n}}{ \frac{2^n-f(n)}{2^n} \frac{1}{2^n}}+ \frac{2^n-f(n)}{2^n} \frac{1}{2^n} \log \frac{\frac{1}{2^n}}{\frac{1}{2^n}(2-\frac{f(n)}{2^n})}             \nonumber  \\
%&&+ \frac{f(n)}{2^n} [n-\log(2- \frac{f(n)}{2^n})]      \nonumber  \\
&&=-\frac{[2^n-f(n)]^2}{2^{2n}} \log(1- \frac{f(n)}{2^n})-\frac{(2^n-f(n))f(n)}{2^{2n}}\log (2- \frac{f(n)}{2^n}) \nonumber  \\
&&+ \frac{f(n)}{2^n} [n-\log(2- \frac{f(n)}{2^n}) ] .  \nonumber  \\
&&=\frac{f(n)n}{2^n}-\frac{[2^n-f(n)]^2}{2^{2n}} \log(1- \frac{f(n)}{2^n})-\frac{(2^n-f(n))f(n)+2^{n}f(n)}{2^{2n}}\log (2- \frac{f(n)}{2^n})\nonumber   \\
&&=\frac{f(n)n}{2^n}-\frac{[2^n-f(n)]^2}{2^{2n}} \log(1- \frac{f(n)}{2^n})-\frac{(2^{n+1}-f(n))f(n)}{2^{2n}}\log (2- \frac{f(n)}{2^n})\nonumber
\end{eqnarray}
Thus we have that  $\lim_{n \rightarrow \infty} \max_{\Theta} I(M;{\bf Y}(\Theta))=0$ and therefore the code is strongly secure, if $\lim_{n\rightarrow \infty} \frac{nf(n)}{2^n} =0$.

Next, we consider the semantic security of this code. For the worst case, we assume the eavesdropper not only knows the code but also the probability distribution of the messages. To show it is not semantically secure, it is sufficient to assume the random message/input ${\bf X}$ has a distribution such that $Pr({\bf X}={\bf 0})=\frac{1}{2}$ and $Pr({\bf X}={\bf x})=\frac{1}{2(2^n-1)}, {\bf x} \in \{0,1\}^n\setminus \{{\bf 0}\}$, where ${\bf 0}$ is the all-zero sequence $(0,0,\dots,0)$. Then a good strategy for the eavesdropper is to access $V_{\Theta}$ for a $\Theta$ containing ${\bf 0}$. In this case, by(\ref{eq_01}) we have that
\begin{eqnarray}
&&Pr({\bf Y}(\Theta))={\bf 0}))=\sum_{{\bf x}}Pr({\bf X}={\bf x}) V_{\Theta}({\bf 0}|{\bf x})  \nonumber \\
&&=\sum_{{\bf x} \not\in \Theta}Pr({\bf X}={\bf x}) V_{\Theta}({\bf 0}|{\bf x})+Pr({\bf X}={\bf 0}) V_{\Theta}({\bf 0}|{\bf 0})  \nonumber \\
&&=\frac{2^n-f(n)}{2(2^n-1)}\frac{1}{2^n}+\frac{1}{2}=\frac{1}{2}[1+\frac{2^n-f(n)}{2^n(2^n-1)}].   \nonumber
 \end{eqnarray}
Because $\sum_{\bf y} V_{\Theta}({\bf y}|{\bf x})  \log \frac{ V_{\Theta}({\bf y}|{\bf x}) }{Pr(Y(\Theta)={\bf y})} \ge 0$, for all ${\bf x}$ and ${\bf 0} \in \Theta$, by (\ref{eq_01}), we have that
\begin{eqnarray}
&&I(M;{\bf Y}(\Theta))=\sum_{{\bf x}}\sum_{\bf y} Pr({\bf X}={\bf x}) V_{\Theta}({\bf y}|{\bf x})  \log \frac{ V_{\Theta}({\bf y}|{\bf x}) }{Pr(Y(\Theta)={\bf y})} \nonumber   \\
&&\ge \sum_{\bf y} Pr({\bf X}={\bf 0}) V_{\Theta}({\bf y}|{\bf 0})  \log \frac{ V_{\Theta}({\bf y}|{\bf 0}) }{Pr(Y(\Theta)={\bf y})}   \nonumber    \\
&&=Pr({\bf X}={\bf 0}) V_{\Theta}({\bf 0}|{\bf 0})  \log \frac{ V_{\Theta}({\bf 0}|{\bf 0}) }{Pr(Y(\Theta)={\bf 0})}   \nonumber    \\
&&=\frac{1}{2}\log \frac{1}{\frac{1}{2}[1+\frac{2^n-f(n)}{2^n(2^n-1)}]} =\frac{1}{2}\log \frac{2}{1+\frac{2^n-f(n)}{2^n(2^n-1)}}  \rightarrow \frac{1}{2},  \nonumber
\end{eqnarray}
as $n \rightarrow \infty$. That is, the code is not semantically secure.

Let $f(n)=2^{na}$ for $0<a<1$. Then, in general, we can show that there is no weak semantic code for the channel having a rate $r > 1-a$ and for all $r<1-a$, there is a semantically secure code with rate $r$, as follows.

{\it Case 1: rate $r > 1-a$,} Suppose that we are given a decodable code with  the message set $\{1,2,\ldots, 2^{nr}\}$ with $r > 1-a$. Let $b=1-r$ and then $b < a$.% and probability of error $\lambda_n$.
 Let  $\Phi(\cdot|i), i=1,2, \ldots, 2^{nr}$ 
be random encoders and ${\cal D}_i, i=1,2, \ldots 2^{nr}$  be decoding sets for the code, respectively.  
Then the decoding sets ${\cal D}_j, j=1,2,\ldots, 2^{nr}$ partition the output space into $2^{nr}$ parts. W. l.o.g, (by renaming the messages) we assume that $|{\cal D}_i| \le |{\cal D}_j|$, if $i \le j$. Then we have that 
\[|\bigcup_{i=1}^{2^{n(a-b)}}{\cal D}_i | \le 2^{n(a-b)} \frac{2^n}{2^{nr}}=2^{na}.   \]
Let $M$ be a random message taking values in $\{1,2,\ldots, 2^{nr}\}$ with probability distribution:
\begin{equation} \label{eq_02}
Pr\{M=i\}=\left\{\begin{array}{ll}
\frac{1-g(n)}{2^{n(a-b)}}  & i \le 2^{n(a-b)}  \\ 
\frac{g(n)}{2^n-2^{n(a-b)}} & i >2^{n(a-b)}  
\end{array} \right.
\end{equation}
for a function $g$ of $n$ such that $g(n) \rightarrow 0$ as $n \rightarrow  \infty$. At this situation, the best way to have information for the eavesdropper is to take a $2^{na}$-subset $\Theta$ covering $\bigcup_{i=1}^{2^{n(a-b)}}{\cal D}_i $ and we assume it is in the case. For $ i \le 2^{n(a-b)}$ and a ${\bf x}$ in the support set of $\Phi(\cdot|i)$, as $W^{G,n}$ is noiseless, $W^{G,n}({\cal D}_i|{\bf x})=0$ if ${\bf x} \not\in {\cal D}_i$ and by (\ref{eq_01}) $W^{G,n}({\cal D}_i|{\bf x})=V_{\Theta}({\cal D}_i|{\bf x})=1$, if ${\bf x} \in {\cal D}_i \subset    \Theta$. Thus, $\sum_{{\bf x}}\Phi({\bf x}|i)W^{G,n}({\cal D}_i|{\bf x})\le\sum_{{\bf x}}\Phi({\bf x}|i)V_{\Theta}({\cal D}_i|{\bf x})$. Consequently, by accessing the channel $V_{\Theta}$ with $  \bigcup_{i=1}^{2^{n(a-b)}}{\cal D}_i  \subset    \Theta$ and using ${\cal D}_i, i=1,2, \ldots 2^{nr}$ to decode, the eavesdropper can decode the message with average probability of error, with respect to the probability defined in (\ref{eq_02}), no larger than
\begin{eqnarray}
&&1-\sum_{i=1}^{2^{nr}}Pr\{M=i\}\sum_{{\bf x}}\Phi({\bf x}|i)V_{\Theta}({\cal D}_i|{\bf x}) \nonumber   \\
&&=1-[\sum_{i=1}^{2^{n(a-b)}}Pr\{M=i\}\sum_{{\bf x}}\Phi({\bf x}|i)V_{\Theta}({\cal D}_i|{\bf x})+\sum_{i=2^{n(a-b)}+1}^{2^{nr}}Pr\{M=i\}\sum_{{\bf x}}\Phi({\bf x}|i)V_{\Theta}({\cal D}_i|{\bf x})]
\nonumber   \\
&& \le 1-\sum_{i=1}^{2^{n(a-b)}}Pr\{M=i\}\sum_{{\bf x}}\Phi({\bf x}|i)V_{\Theta}({\cal D}_i|{\bf x})%+\sum_{i=2^{n(a-b)}+1}^{2^{nr}}Pr\{M=i\}]
 \nonumber   \\
%&&= 1-[\sum_{i=1}^{2^{n(a-b)}}Pr\{M=i\}\sum_{{\bf x}}\Phi({\bf x}|i)V_{\Theta}({\cal D}_i|{\bf x})+\g(n)]               \nonumber   \\
%&&=1-[\sum_{i=1}^{2^{n(a-b)}}Pr\{M=i\}\sum_{{\bf x}}\Phi({\bf x}|i)V_{\Theta}({\cal D}_i|{\bf x})+g(n)] \nonumber   \\
&&  \le 1-\sum_{i=1}^{2^{n(a-b)}}Pr\{M=i\}\sum_{{\bf x}}\Phi({\bf x}|i)W^{G,n}({\cal D}_i|{\bf x}) \nonumber   \\
&&-\sum_{i=2^{n(a-b)}+1}^{2^{nr}}Pr\{M=i\}\sum_{{\bf x}}\Phi({\bf x}|i)W^{G,n}({\cal D}_i|{\bf x})+\sum_{i=2^{n(a-b)}+1}^{2^{nr}}Pr\{M=i\} \nonumber   \\
&&= 1-\sum_{i=1}^{2^{nr}}Pr\{M=i\}\sum_{{\bf x}}\Phi({\bf x}|i)W^{G,n}({\cal D}_i|{\bf x})+g(n)   \rightarrow 0 \nonumber 
\end{eqnarray}
as $n\rightarrow \infty$, whenever the maximum or average probability with respect to the probability defined in (\ref{eq_02}) of the code for the main channel $W^{G,n}$ goes to $0$ as $n$ increases. That is, the eavesdropper can decode correctly with a probability asymptotically approaching one! Thus, in a standard way, by Fano inequality, we conclude that for a $h(n)$ with $h(n) \rightarrow 0$,  $n \rightarrow 0$
\begin{eqnarray}
&&\frac{1}{n} I(M;{\bf Y}(\Theta))\ge \frac{1}{n}H(M)-h(n) \ge -\frac{1}{n}\sum_{i=1}^{2^{n(a-b)}}Pr\{M=i\}\log Pr\{M=i\}-h(n) \nonumber   \\
&&=\frac{1-g(n)}{n} \log \frac{2^{n(a-b)}}{1-g(n)}-h(n)   \rightarrow a-b>0   \nonumber
\end{eqnarray}
as $n \rightarrow 0$. Namely, the code is not weakly semantically secure.

{\it Case 2: $r <1-a,$} Now, we suppose $r:=1-b < 1-a$, i.e., $a<b$. Now we have to construct a semantically secure code, with rate $r$. We first partition $\{0,1\}^n$ into $2^{nr}$ subsets ${\cal A}_i,i=1,2,\ldots,2^{nr}$ of equal size $2^{n(1-r)}=2^{nb}$. We let the random encoder for message $i$, $\Phi(\cdot|i)$ be a uniform distribution on ${\cal A}_i$ and the decoding set for message $i$ on the main channel ${\cal D}_i={\cal A}_i$ for $i=1,2,\ldots,2^{nr}$. Then obviously,  $\sum_{{\bf x} \in {\cal A}_i}\Phi({\bf x}|i)W^{G,n}({\cal D}_i|{\bf x})=1$ and this is a zero error code for the legitimate receiver, and we only need to show the code is also semantically secure. To this end, we fix an arbitrary $\Theta \in {\{0,1\}^n \choose 2^{na}}$ and calculate the conditional distribution of output $\sum_{{\bf x} \in {\cal A}_i}\Phi({\bf x}|i)V_{\Theta}(\cdot|{\bf x})$ under the condition given a message $i \in \{1,2,\ldots,2^{nr}\}$.

For ${\bf y} \in {\cal A}_i \cap \Theta$, we have that by (\ref{eq_01})
\begin{eqnarray}
&&\sum_{{\bf x} \in {\cal A}_i}\Phi({\bf x}|i)V_{\Theta}({\bf y}|{\bf x})=\sum_{{\bf x} \in {\cal A}_i\cap \Theta}\Phi({\bf x}|i)V_{\Theta}({\bf y}|{\bf x})+\sum_{{\bf x} \in {\cal A}_i \setminus \Theta}\Phi({\bf x}|i)V_{\Theta}({\bf y}|{\bf x})   \nonumber      \\
&&=\frac{1}{2^{nb}} +\sum_{{\bf x} \in {\cal A}_i \setminus \Theta}\Phi({\bf x}|i)V_{\Theta}({\bf y}|{\bf x})=  \frac{1}{2^{nb}}+\frac{|  {\cal A}_i \setminus \Theta|}{2^{nb}} \frac{1}{2^n}=  \frac{1}{2^{nb}}[1+\frac{|  {\cal A}_i \setminus \Theta|}{2^{n}}].       \nonumber     
\end{eqnarray}
For ${\bf y} \not\in \Theta$, we have that by (\ref{eq_01})
\begin{eqnarray}
&&\sum_{{\bf x} \in {\cal A}_i}\Phi({\bf x}|i)V_{\Theta}({\bf y}|{\bf x})=\sum_{{\bf x} \in {\cal A}_i\cap \Theta}\Phi({\bf x}|i)V_{\Theta}({\bf y}|{\bf x})+\sum_{{\bf x} \in {\cal A}_i \setminus \Theta}\Phi({\bf x}|i)V_{\Theta}({\bf y}|{\bf x})   \nonumber      \\
&&=\sum_{{\bf x} \in {\cal A}_i \setminus \Theta}\Phi({\bf x}|i)V_{\Theta}({\bf y}|{\bf x})=  \frac{|  {\cal A}_i \setminus \Theta|}{2^{nb}} \frac{1}{2^n}=  \frac{|  {\cal A}_i \setminus \Theta|}{2^{n(1+b)}}  .      \nonumber     
\end{eqnarray}
For ${\bf y} \in \Theta \setminus {\cal A}_i$ by (\ref{eq_01}), we have that
%\begin{equation}
%\sum_{{\bf x} \in {\cal A}_i}\Phi({\bf x}|i)V_{\Theta}({\bf y}|{\bf x})=0. \nonumber     
%\end{equation}
\begin{equation}
\sum_{{\bf x} \in {\cal A}_i}\Phi({\bf x}|i)V_{\Theta}({\bf y}|{\bf x})=\sum_{{\bf x} \in {\cal A}_i \setminus \Theta}\Phi({\bf x}|i)V_{\Theta}({\bf y}|{\bf x}) =  \frac{|  {\cal A}_i \setminus \Theta|}{2^{nb}} \frac{1}{2^n}=  \frac{|  {\cal A}_i \setminus \Theta|}{2^{n(1+b)}}  .  \nonumber    
\end{equation}

Hence, for $i=1,2,\ldots,2^{nr}$, and the uniform distribution $Q_U$ on $\{0,1\}^n$, 
\begin{eqnarray} \label{eq_03}
&& \sum_{{\bf y} \in \{0,1\}^n} [\sum_{{\bf x} \in {\cal A}_i}\Phi({\bf x}|i)V_{\Theta}({\bf y}|{\bf x})] \log \frac{[\sum_{{\bf x} \in {\cal A}_i}\Phi({\bf x}|i)V_{\Theta}({\bf y}|{\bf x})]}{Q_U({\bf y}) }           \nonumber     \\
&&= \sum_{{\bf y} \in  {\cal A}_i \cap \Theta} [\sum_{{\bf x} \in {\cal A}_i}\Phi({\bf x}|i)V_{\Theta}({\bf y}|{\bf x})] \log \frac{[\sum_{{\bf x} \in {\cal A}_i}\Phi({\bf x}|i)V_{\Theta}({\bf y}|{\bf x})]}{Q_U({\bf y}) }    \nonumber     \\
&&+ \sum_{{\bf y}  \not\in \Theta} [\sum_{{\bf x} \in {\cal A}_i}\Phi({\bf x}|i)V_{\Theta}({\bf y}|{\bf x})] \log \frac{[\sum_{{\bf x} \in {\cal A}_i}\Phi({\bf x}|i)V_{\Theta}({\bf y}|{\bf x})]}{Q_U({\bf y}) } \nonumber     \\
&&+ \sum_{{\bf y} \in \Theta \setminus {\cal A}_i} [\sum_{{\bf x} \in {\cal A}_i}\Phi({\bf x}|i)V_{\Theta}({\bf y}|{\bf x})] \log \frac{[\sum_{{\bf x} \in {\cal A}_i}\Phi({\bf x}|i)V_{\Theta}({\bf y}|{\bf x})]}{Q_U({\bf y}) } \nonumber     \\
&&= \sum_{{\bf y} \in  {\cal A}_i \cap \Theta}  \frac{1}{2^{nb}}[1+\frac{|  {\cal A}_i \setminus \Theta|}{2^{n}} ] \log \frac{\frac{1}{2^{nb}}[1+\frac{|  {\cal A}_i \setminus \Theta|}{2^{n}}]}{Q_U({\bf y}) }  + \sum_{{\bf y}  \not\in \Theta} [\frac{|  {\cal A}_i \setminus \Theta|}{2^{n(1+b)}} ] \log \frac{\frac{|  {\cal A}_i \setminus \Theta|}{2^{n(1+b)}}}{Q_U({\bf y}) }  \nonumber     \\
%&&= \sum_{{\bf y} \in  {\cal A}_i \cap \Theta}  \frac{1}{2^{nb}}[1+\frac{|  {\cal A}_i \setminus \Theta|}{2^{n}} ] \log \frac{[\sum_{{\bf x} \in {\cal A}_i}\Phi({\bf x}|i)V_{\Theta}({\bf y}|{\bf x})]}{Q_U({\bf y}) }  + \sum_{{\bf y}  \not\in \Theta} [\frac{|  {\cal A}_i \setminus \Theta|}{2^{n(1+b)}} ] \log \frac{\frac{|  {\cal A}_i \setminus \Theta|}{2^{n(1+b)}}}{Q_U({\bf y}) }  \nonumber     \\
&&+ \sum_{{\bf y}  \in \Theta \setminus {\cal A}_i} [\frac{|  {\cal A}_i \setminus \Theta|}{2^{n(1+b)}} ] \log \frac{\frac{|  {\cal A}_i \setminus \Theta|}{2^{n(1+b)}}}{Q_U({\bf y}) }  \nonumber     \\
\end{eqnarray}

As $Q_U({\bf y})=\frac{1}{2^n}, |{\cal A}_i|=2^{nb}$, and $|\Theta|=2^{na}, a<b$, the first term of $(\ref{eq_03})$ is upper bounded by
%\[\frac{|{\cal A}_i \cap \Theta|}{2^{nb}}[1+\frac{|  {\cal A}_i| -|\Theta|}{2^{n}} ] \log \frac{1}{Q_U({\bf y}) } \le 2^{-n(b-a)} (1+2^{n(1-b)})n     
\begin{eqnarray}
&&\frac{|{\cal A}_i \cap \Theta|}{2^{nb}}[1+\frac{|  {\cal A}_i\setminus\Theta|}{2^{n}} ] \log 2^{n(1-b)}[1+\frac{|  {\cal A}_i\setminus\Theta|}{2^{n}} ]               \nonumber                       \\
&& \le \frac{|\Theta|}{2^{nb}}[1+\frac{|  {\cal A}_i| }{2^{n}} ] \log  2^{n(1-b)}[1+\frac{|  {\cal A}_i|}{2^{n}} ]               \nonumber                       \\   
&&=2^{-n(b-a)}[1+2^{-n(1-b)}]\log   2^{n(1-b)} [1+2^{-n(1-b)}]                \nonumber  
\end{eqnarray}
and the second term is %upper bounded by
\begin{equation}
 \sum_{{\bf y}  \not\in \Theta} [\frac{|  {\cal A}_i\setminus \Theta|}{2^{n(1+b)}} ] \log \frac{|  {\cal A}_i \setminus \Theta|}{2^{nb}} \le 0 .               \nonumber     
\end{equation}
Similarly, the last term is non-positive as well.
%\[ n \sum_{{\bf y}  \not\in \Theta} \frac{|  {\cal A}_i \setminus \Theta|}{2^{n(1+b)}}  \log \frac{|  {\cal A}_i \setminus \Theta|}{2^{n(1+b)}}  \le 0. \]
Thus, by the Information Radius Lemma due to  I. Csisz\'{a}r, we conclude that
\[I(M;{\bf Y}(\Theta) ) \le 2^{-n(b-a)}[1+2^{-n(1-b)}]\log   2^{n(1-b)} [1+2^{-n(1-b)}]    \rightarrow 0 \]
as $n$ increases, for all random message $M$ taking values in $\{1,2,\ldots, 2^{nr}\}$. Indeed, the code is semantically secure.

\section{proof of theorem \ref{the: question 3}}\label{sec: proof of theorem question 3}
We have seen an example in the proof of Theorem \ref{the: question 2}, where $|{\cal W}^{G,n}|=1$ (by the  Convention, $\log \log |{\cal W}^{G,n}|:=0$), $\log \log v_n^{-1}=\log n$, and $\lim_{n \rightarrow \infty}\frac{1}{n} \log \log |{\cal  V}^{G,n}|=a$ and $C_{a-str}({\cal W}^{G,n},{\cal V}^{G,n})=1,C_{m-mis}({\cal W}^{G,n},{\cal V}^{G,n})=1-a$ i.e., the bound in (\ref{eq_thm2}) is tight. So it is sufficient for us to show  (\ref{eq_thm2}).

Denote by \[a:= \max \{\limsup_{n \rightarrow \infty}  \frac{1}{n}  \log \log |{\cal  W}^{G,n}|,\limsup_{n \rightarrow \infty}  \frac{1}{n}   \log \log |{\cal  V}^{G,n}| \}.\]

For all $\epsilon, \lambda, \mu >0$, for sufficiently large $n$, we have that
\begin{equation} \label{eq_gavwc1}
\log \log |{\cal  W}^{G,n}| < n(a+\frac{\epsilon}{4}) \mbox{ and } \log \log |{\cal  V}^{G,n}| < n(a+\frac{\epsilon}{4}),
\end{equation}
and there exists an average error and strongly secure code $(\Phi_n, \psi_n)$ with message set ${\cal U}^{G,n}$ and rate $\frac{1}{n} \log {|{\cal U}^{G,n}|} >C_{a-str}(({\cal W}^{G,n},{\cal V}^{G,n}))-\frac{\epsilon}{4}$ such that for uniform random message $U$ on ${\cal U}^{G,n}$, (\ref{eq_ex01avc}) and (\ref{eq_ex02avc}) hold. Again, by (\ref{eq_dv}) we may use $\log v_n^{-1}$ as $\delta_n$ in Lemma \ref{lemma_semaim}. To apply Lemma \ref{lemma_semaim}, we let $K_n=2^{n(a+\frac{\epsilon}{2})}, \beta_n=\frac{1}{n}$, such that
\[ \frac{1}{n}\log J_n=\frac{1}{n}[\log|{\cal U}^{G,n}|- n(a+\frac{\epsilon}{2})-\log n]>C_{a-str}(({\cal W}^{G,n},{\cal V}^{G,n})-a-\epsilon,\]
for sufficiently large $n$. As in the proof of Theorem \ref{thmavwc},  we choose $A_n=\frac{1}{2}K_n \mu$ and $B_n=\frac{1}{2}K_n\lambda$ and then (\ref{eq_mlmm01}) holds after replacing $\mu$ and $\lambda$ by $\frac{\mu}{8} $ and $ \frac{\lambda}{8}$ respectively, in (\ref{eq_mlmm01}). Then to continue our proof, we have to verify (\ref{eq_mlmm02}). In this case, we have that by ({\ref{eq_gavwc1})
\begin{eqnarray}
&&\log |{\cal V}^{G,n}|e^{-\frac{\delta_n^{-1}A_n}{4}}=2^{\log \log  |{\cal V}^{G,n}|}-\frac{\delta_n^{-1}A_n}{4}\log e   \nonumber                       \\
&&< 2^{ n(a+\frac{\epsilon}{4})}-  \frac{\delta_n^{-1}K_n\mu}{8}  \log e \nonumber                       \\
&&= 2^{ n(a+\frac{\epsilon}{4})}-  \frac{2^{n(a+\frac{\epsilon}{2})}\mu}{8\log v_n^{-1}}  \log e  \nonumber \\
&&= 2^{ n(a+\frac{\epsilon}{4})}[1-\frac{2^{n(\frac{\epsilon}{4}-\frac{1}{n} \log \log v_n^{-1})}\mu}{8}  \log e] <- 2^{ n(a+\frac{\epsilon}{4})}, \nonumber
\end{eqnarray}
if $n$ is sufficiently large, which yields the first inequality in (\ref{eq_mlmm02}), because $\log J_n$ is linearly increasing as $n$ grows. Similarly, we can prove  the second inequality in (\ref{eq_mlmm02}). Finally, the third inequality in (\ref{eq_mlmm02}) can be shown in the same way as in the proof of Theorem \ref{thmavwc}. (In fact, in the current proof,  ``$J_n$" is smaller than that in the previous proof and ``$K_n$" is larger than that in the previous proof, and so the left-hand side of the inequality is even smaller than that in the previous proof). Thus,  by Lemma \ref{lemma_semaim}, we obtain a  maximum error and semantically secure code $(\Upsilon_n, \nu_n)$ with a rate larger than $C_{a-str}(({\cal W}^{G,n},{\cal V}^{G,n})-a-\epsilon$ such that  (\ref{eq_mlmm03}) and (\ref{eq_mlmm04}) hold. By substitution of $A_n=\frac{1}{2}K_n \mu$ and $B_n=\frac{1}{2}K_n\lambda$ in (\ref{eq_mlmm03}) and (\ref{eq_mlmm04}), we obtain that
 $W^{G,n}(\nu_n({\bf Y})\not=j|\Upsilon_n(j)) < \lambda,$ and $I(\tilde{U};\tilde{\bf Z}(V^{G,n}))< \mu$. Hence, $C_{a-str}(({\cal W}^{G,n},{\cal V}^{G,n})-a$ is achievable by maximum error and semantically secure codes. That is, (\ref{eq_thm2}) and our proof is completed.

\section*{Acknowledges}
Holger Boche would like to thank Dr. Manfred Lochter of the Federal Office for Information Security (BSI) for his numerous questions regarding the identification of gaps in the arguments surrounding security discussions on QKD, which served as the starting point for this work. These questions focused in particular on practically relevant security measures beyond strong security and the impact of active attacks. This paper answers these questions and would not have been possible without Dr. Manfred Lochter’s continuous scrutiny of the security issues of QKD.
\ifCLASSOPTIONcaptionsoff
  \newpage
\fi

\bibliographystyle{ieeetr} 
\bibliography{ref}
\end{document}